\documentclass[prd,twocolumn,superscriptaddress,floatfix,amsmath,amssymb,nofootinbib]{revtex4-1}

\usepackage[utf8]{inputenc}
\usepackage{amsmath}
\usepackage{amssymb}
\usepackage{physics}
\usepackage{ifthen}
\usepackage{xcolor}
\usepackage[normalem]{ulem}
\usepackage{mathtools}
\usepackage{graphicx}
\usepackage{natbib}
\usepackage{braket}
\usepackage{url}
\usepackage{hyperref}
\usepackage{bm}

\usepackage{soul}
\usepackage{enumerate}

\def\shownal{1}
\newcommand{\nemma}[1]{\ifthenelse{\shownal=1}{\textcolor{purple}{[[Emma: #1]]}}{}}

\newcommand{\ii}{\mathrm{i}}
\renewcommand{\dd}{\mathrm{d}}

\begin{document}

% ------------------------------------ TITLE
\title{Work distributions on quantum fields}

% author order not yet agreed upon!

\author{Alvaro Ortega}
\affiliation{Department of Applied Mathematics, University of Waterloo, Waterloo, Ontario, N2L 3G1, Canada}
\author{Emma McKay}
\affiliation{Department of Applied Mathematics, University of Waterloo, Waterloo, Ontario, N2L 3G1, Canada}
\affiliation{Institute for Quantum Computing, University of Waterloo, Waterloo, Ontario, N2L 3G1, Canada}
\author{\'Alvaro M. Alhambra}
\affiliation{Perimeter Institute for Theoretical Physics, 31 Caroline St. N., Waterloo, Ontario, N2L 2Y5, Canada}
\author{Eduardo Mart\'{i}n-Mart\'{i}nez}
\affiliation{Department of Applied Mathematics, University of Waterloo, Waterloo, Ontario, N2L 3G1, Canada}
\affiliation{Institute for Quantum Computing, University of Waterloo, Waterloo, Ontario, N2L 3G1, Canada}
\affiliation{Perimeter Institute for Theoretical Physics, 31 Caroline St. N., Waterloo, Ontario, N2L 2Y5, Canada} 
% ------------------------------------------ ABSTRACT
\begin{abstract}
    We study the work cost of processes in quantum fields without the need of projective measurements, which are always ill-defined in quantum field theory. Inspired by interferometry schemes, we propose a work distribution that generalizes the two-point measurement scheme employed in quantum thermodynamics to the case of quantum fields and avoids the use of projective measurements. The distribution is calculated for local unitary processes performed on KMS (thermal) states of scalar fields. Crooks theorem and the Jarzynski equality are shown to be satisfied for a family of spatio-temporally localized  unitaries, and some features of the resulting distributions are studied as functions of temperature and the degree of  localization of the unitary operation. %\nemma{as the abstract is quite long, I suggest removing one or both of the last two sentences.} %We show that the average work cost of a local unitary process is independent of temperature, while the variance monotonically increases with it.
    We show how the work fluctuations become much larger than the average as the process becomes more localized in both time and space.
\end{abstract}

\maketitle

% ------------------------------------------ INTRO

\textit{\textbf{Introduction.-}}  At microscopic scales average quantities no longer characterize completely the state of a system or the features of a thermodynamic process. There, stochastic or quantum fluctuations become relevant, being of the same order of magnitude as the expectation values \cite{campisicol,esposito,jarz2}. It is therefore important to develop tools that allow us to study the properties of these fluctuations to fully understand thermodynamics at the small scales.

One of the best studied quantities in this context is work of out-of-equilibrium processes, and its associated fluctuations. The notion of work is an empirical cornerstone of macroscopic equilibrium thermodynamics. However, work in microscopic quantum scenarios is a notoriously subtle concept (e.g., it cannot be associated to an observable \cite{talkner}), and although there is no single definition of work distributions and work fluctuations in quantum theory, several possibilities have been proposed (see   e.g., \cite{Baeumer2018}). Perhaps the most established notion of work fluctuations is that defined through the Two-Point  Measurement (TPM) scheme \cite{tasaki2000jarzynski, referee}, where the work distribution of a  process is obtained by performing two projective measurements of the system's energy, at the beginning and at the end of the process. The TPM formalism defines a work distribution  with a number of desirable properties: it is linear on the input states, it agrees with the unambiguous classical definition for states diagonal in the energy eigenbasis, and it yields a number of fluctuation theorems in different contexts \cite{campisicol,espositofluct, referee}.
%\st{For instance, it satisfies the Jarzynski equality %\cite{campisifluct,espositofluct,jorge,hal,talkner,campisifluc2}
%, it can be used to study both closed and open systems 
%\cite{Campisi2009}
%, and it matches the classical theory when the states do not have coherence.} %\cite{Jarz2015, zhu2016}.}
%, and the predictions of some of them have been tested experimentally \cite{batalhaoexp}.}

An important caveat of this definition is that it cannot be straightforwardly generalized to processes involving quantum fields: projective measurements in quantum field theory (QFT) are incompatible with its relativistic nature.   They cannot be localized \cite{Redhead1995}, they can introduce ill-defined operations due to UV divergences and, among other serious problems, they enable superluminal signaling even in the most innocent scenarios \cite{sorkin}. 
% \nemma{I think it would be really helpful for thermo people to have some explicit discussion of the problems with energy eigenstates: that they are nonlocal, that local projections onto finite subsets of eigenstates are impossible.} {\color{magenta} \bf [EDU: while I agree in spirit, we are a bit short on space, and I believe that the references do explain that, even in a pedagogical way in Fay's ``Useless qubits" paper. I have, hoever, included a bit of this along the lines of your comment in the conclusion]} 
For these reasons, it has been strongly argued that projective measurements should be banished from the formalism of any relativistic field theory \cite{sorkin,Dowker,Dowker2}. However, quantum fields are certainly subject to a wealth of thermodynamic and non-equilibrium phenomena, and as such it should be possible to define an operationally meaningful work distribution, potentially different from the standard TPM scheme. One avenue to build such a work distribution is through the ability to operate on quantum fields through locally coupling other systems, such as e.g., atoms or particle detectors. This allows the performance of  measurements on the field that are well-defined \cite{Fewster}  and physically meaningful \cite{Martin-Martinez2018}. Thus, whichever definition we construct for the work distribution, it should be based on such physically attainable localized measurements, and should not rely on projective measurements as  previous works attempted (e.g., \cite{bartolotta2018jarzynski}).

In recent works \cite{Dorner2013,Mazzola2014}, it was shown that the complete work distribution given by TPM scheme for a finite dimensional system can be  measured by performing measurements on an auxiliary qubit, in what is called a Ramsey interferometric scheme. %That is, in finite-dimensional systems, in which the TPM scheme is well-defined, the Ramsey scheme gives an equivalent characterization of the work fluctuations \nemma{are you using the word fluctuations to mean work statistics or its second moment? also this second sentence sounds very much like the first} \alvo{I use work fluctuation as a synonym of work probability distribution. If somewhere the meanng is diferent, then it is incorrect}.
This was experimentally implemented in \cite{batalhaoexp}. Inspired by  this idea, we propose a definition of  work distributions in quantum fields based on the Ramsey scheme which is in fact well defined for a QFT despite the impossibility of projective measurements.  We show that this new distribution satisfies the usual Jarzynski and Crooks theorems when the field is initially in a KMS state (the states that generalize thermal Gibbs states for quantum fields \cite{Kubo, schwinger}) and evolves through a spatially localized unitary. This shows that such work distribution is well-defined for fields even though projective measurements are not. We also obtain analytical expressions for the variance and the average of the work distribution for some useful simple cases of local field operations.  Finally we discuss how, through either Crooks or Jarzynski's theorems, the proposed \st{work} distribution can be used as a new way of computing ratios of partition functions between field theories,  potentially yielding simpler approaches to the problem than  path integral methods.

%The paper is organized as follows. In section \ref{sec:ramsey}, we review the Ramsey scheme for work distributions, and argue for its extension to quantum fields. In section \ref{sec:calculations}, we obtain the work characteristic function for KMS states of scalar fields, and the analytical form of the work distribution for spherically symmetric smearings. In section \ref{sec:analysis} we comment on some of the statistical properties of the obtained distributions, and we verify that the Jarzynski inequality is satisfied. The details of some of the calculations involved in \ref{sec:calculations} can be found in \ref{app:vacuum} and \ref{app:kms}

% ------------------------------------------ RAMSEY SCHEME
    
\textit{\textbf{TPM work distributions and Ramsey scheme.-}}   Consider a quantum field initially in an equilibrium KMS state $\hat \rho$ of temperature $\beta^{-1}$, which is driven out of equilibrium by a time-dependent Hamiltonian $\hat H(t)$, turned on during an interval $[0,T]$. The work distribution quantifies the work cost of the unitary process on the field $\hat U(T,0)$ generated by the Hamiltonian $\hat H(t)$.

 As discussed above  projective measurements cannot be implemented in quantum fields because they are incompatible with relativistic causality  \cite{sorkin,Dowker,Dowker2}. Thus, the TPM scheme cannot be readily applied  to processes involving quantum fields. However, the Ramsey scheme, which only involves interactions with a low-dimensional ancilla, provides an indirect way to gather the same work statistics.
%\alvo{If space is an issue, maybe remove from here unil the definition of reversed process?} 
For completeness, let us review the TPM scheme to define a work distribution. The steps are the following:

%Ramsey interferometry has been used in other contexts, such as parameter estimation \cite{interf1,dechiaraint}, prior to its application to measuring the work distribution of a process occurring on a finite dimensional quantum system as presented in \cite{Dorner2013,Mazzola2014}. More precisely, the work distribution prescribed by the TPM scheme is obtained through this scheme. Ideal projective measurements can't be implemented in quantum fields, since it would allow two spacelike separated observers to signal each other \cite{sorkin,Dowker}. Thus, the TPM scheme cannot be readily generalized to processes involving quantum fields. One might suggest that a desirable property of any definition of work distribution is that it can be applied to any quantum system of interest. In this sense, the Ramsey scheme provides an alternative way to define work distributions that can be used for quantum fields, with a spirit similar to the one introduced in \cite{Fewster}. Properties of quantum fields are measured through the response of physical systems coupled to them. This definition has a nice limit when the system becomes finite dimensional, in the sense that it predicts the same work distribution as the TPM scheme, which has had many successes up to date. The steps of the TPM scheme are the following: 

\begin{enumerate}
    \item A projective measurement of $\hat H(0)$ is done on the initial state $\hat\rho$. This yields the energy measured as $E_i$ and the post-measurement state $\ket{E_i}\!\bra{E_i}$.
    \item Unitary evolution of the post-measurement state according to the unitary associated to the process $\hat U(T,0)$.
    \item  A projective measurement of $\hat H(T)$ is done on $\hat U(T,0)\ket{E_i}\!\bra{E_i}\hat U^{\dagger}(T,0)$, returning the value $E_j'$.
\end{enumerate}

The  possible values of the work $w^{(ij)}$ are defined as $w^{(ij)} = E_j' - E_i$. The work probability distribution is \begin{equation}
    P(W) = \sum_{(ij)} \delta\!\left(W\! -\! w^{(ij)}\right)\bra{E_i}\rho\ket{E_i}|\!\bra{E'_j}\hat U(T,0)\ket{E_i}|^2,
\end{equation}
with a corresponding characteristic function 
\begin{equation} \label{eq:characteristic}
    \widetilde{P}(\mu) = \int P(W) e^{\ii\mu W}dW = \langle e^{i\mu W}\rangle.
\end{equation}

It is also important to define a ``time-reversed" process, in which the driving has the opposite temporal order. That is, 
\begin{enumerate}
	\item A projective measurement is done on the basis of $\hat H(T)$, yielding $E_{j,\text{rev}}'$.
	\item The unitary evolution $\hat U_{\text{rev}}(T,0)$ corresponding to the driven Hamiltonian $\hat H(T-t)$ with $t=[0,T]$ is implemented.
	\item A final projective measurement in the basis of $\hat H(0)$ is implemented returning the value $E_{i,\text{rev}}$.
\end{enumerate}
The corresponding work probability distribution is \begin{align}
P_{\text{rev}}(W) &= \sum_{(ij)} \delta\left(W - w_{\text{rev}}^{(ji)}\right)\\ &\times\bra{E_{j,\text{rev}}'}\rho\ket{E_{j,\text{rev}}'}|\!\bra{E_{i,\text{rev}}}\hat U(T,0)\ket{E_{j,\text{rev}}'}|^2,\nonumber
\end{align}
where $w_{\text{rev}}^{(ji)}=E_{i,\text{rev}}-E_{j,\text{rev}}'$. We can also define $\widetilde{P}_{\text{rev}}(\mu) = \int P_{\text{rev}}(W) e^{\ii\mu W}dW $.

In the original proposals \cite{Dorner2013,Mazzola2014}, Ramsey interferometry was employed to probe the TPM work distributions as follows: the system of interest is coupled to an auxiliary qubit, which engages the system in an evolution conditional on whether the qubit is excited or not. By preparing the qubit in a superposition of ground and excited states, this process transfers the data about the characteristic function of the TPM work distribution to the state of the qubit. This is thus a rather `non-invasive' procedure to acquire statistics which otherwise would require projective measurements. The steps are:
\begin{enumerate}
    \item The system and the auxiliary qubit are prepared in the product state $\hat \rho\otimes\ket{0}\!\bra{0}$, where $\hat \rho$ is the state of the quantum system at the beginning of the thermodynamic process.
    \item A Hadamard gate is applied  on the qubit.
    \item The  system and the auxiliary qubit evolve unitarily according to \begin{equation}
    \label{eq:ram}
       \hat M_{\mu} = \hat U_{S}e^{-\ii\mu \hat H(0)}\otimes\ket{0}\!\bra{0} + e^{-\ii\mu \hat H(T)}\hat U_{S}\otimes\ket{1}\!\bra{1}.
    \end{equation}
    Here $\hat U_{S}$ is the unitary acting on the system between times $0$ and $T$.
    \item A second Hadamard is applied to the qubit.
\end{enumerate}

At the end of this procedure, we obtain that the reduced state of the auxiliary qubit is \mbox{$\hat \rho_{\mu} = \frac{1}{2}\left(\openone + \Re(\widetilde{P}(\mu))\hat\sigma_z + \Im(\widetilde{P}(\mu))\hat\sigma_y \right)$}. By iterating this process over many values of $\mu$ and performing state tomography, the work distribution of any unitary process on a system of interest can then be constructed without projective measurements.

% ------------------------------------------ CALCULATIONS

\textit{\textbf{Work distributions for thermal states of quantum fields.-}}  We will now design a version of the Ramsey scheme to obtain a characteristic function that defines the work distribution of a process, which will be a localized unitary on a scalar field.  Consider a scalar quantum field  $\hat \phi (t,\bm{x})$ written in terms of plane-wave modes as\begin{equation}
    \label{eq:field}
    \hat\phi (t,\bm{x}) = \int \frac{\dd^3 \bm k}{\left(2\pi\right)^{3/2}\sqrt{2\omega_{\bm k}}}\left(\hat a_{\bm{k}}e^{\ii \mathsf{k}\cdot \mathsf{x}} + \hat a^{\dagger}_{\bm{k}}e^{-\ii \mathsf{k}\cdot \mathsf{x}}\right),
\end{equation}
where $\mathsf{k}\cdot \mathsf{x} \coloneqq \bm{k}\cdot \bm{x}-\omega_{\bm{k}}t $, $\omega_{\bm k}=\sqrt{m^2+\bm k^2}$ and $\comm{\hat a_{\bm{p}}}{\hat a^{\dagger}_{\bm{q}}} =\delta^{(3)}\left(\bm{p} - \bm{q}\right)$. We take the field to be in a KMS state \cite{Kubo,schwinger} of inverse temperature $\beta$, $\hat\rho_{\beta}$.  KMS thermality generalizes Gibbs' notion of thermality to cases where, due to the dimensionality of the Hilbert space, Gibbs thermal states are not well-defined. This is the case of QFTs, where usually the partition function is ill-defined. More formally, for a KMS state $\hat\rho_{\beta}$  (with inverse KMS temperature $\beta$) with respect to time translations generated by a Hamiltonian $\hat H$  the two-point correlator $\mathcal{W}_{\hat\rho}(\tau,\tau')\coloneqq\Tr\left[\hat\rho\hat\phi\left(t\left(\tau\right)\bm{x}\left(\tau\right)\right)\hat\phi\left(t\left(\tau'\right)\bm{x}\left(\tau'\right)\right)\right] $ satisfies the following two conditions (see, among many others, \cite{kms,Pipo}):
\begin{enumerate}
    \item $\mathcal{W}_{\rho}(\tau,\tau') = \mathcal{W}_{\rho}(\Delta\tau)$ (Stationarity).
    \item $\mathcal{W}_{\rho}(\Delta\tau + \ii\beta) = \mathcal{W}_{\rho}(-\Delta\tau)$ ($\mathbb{C}-$antiperiodicity). 
\end{enumerate}
% We are interested in studying the work distributions in this family of states since they are the natural generalization of thermal states for quantum fields.
Notice that the vacuum state is a KMS state with \mbox{$\beta \to \infty$}, that  is, zero temperature.

We proceed to characterize the localized unitary we apply on the field. For a free scalar field, any local observable is a linear combination of the field amplitude $\hat\phi$ and its canonical momentum $\hat \pi$. For concreteness, in this letter, we focus on unitaries acting on the field that are generated by Hamiltonians of the form 
% \nemma{why these unitaries? let's add a comment about what they do or at least justify why we are interested in them} {\color{magenta} \bf [EDU: Hi Emma, good point. I was planning to write that in the conlucsions but I added a bit here as well at the beginning of the section and at the end of the pargaph after the equation.]}
\begin{equation} \label{eq:hamiltonian}
\hat H_{\phi}(t) =  \hat H_0+\lambda\chi(t)\int_{\mathbb{R}^3} \dd^3\bm{x} F(\bm{x})\hat \phi(t,\bm{x})=\hat H_0+\hat H_I(t),
\end{equation} 
in the interaction picture, where $\hat H_0$ is the free Hamiltonian of the field, and $\chi(t)$ and $F(\bm{x})$ are the switching and smearing functions, respectively. We assume that the switching function has strong support in a finite region \footnote{or in simple words, there is only a finite time interval where it is not true that $\chi(t)\ll1$. This is the case of compactly supported functions, but also exponentially suppressed functions such as Gaussians.} and, without loss of generality, we take the strong support of the switching function to be in the interval  $[0,T]$, where $0$ and $T$ are the starting and ending times of the process under study. In other words, the field evolves freely (or very approximately freely if the switching function is not strictly compact) except for the interval $[0,T]$ where we perform a spatiotemporally localized  unitary operation on the support of $F(\bm x)$. By doing this, we obtain that $\hat H_{\phi}(0) = \hat H_{\phi}(T) = \hat H_0$, which  simplifies our analysis. This is a particular unitary operation on a localized field observable (representing a multimode displacement operation \cite{ScullyBook}). Considering localized unitaries generated by a smeared $\hat\pi$ is completely analogous, so this particular case is easily generalizable to all localized  unitaries on a free field. %Also, it seems reasonable to restrict our attention to processes localized both in time and space.

%The vacuum state of the field will be denoted as $\ket{\Omega}$, and the ground state of the auxiliary qubit will be denoted $\ket{0}$. Let us formally define what would be a Ramsey scheme on the field.

At the beginning of the Ramsey scheme, the state of the field-qubit system is $
\hat \rho  = \hat\rho_{\beta}\otimes\ket{0}\!\bra{0}$.  Applying the Hadamard on the qubit results in $\hat \rho_0 = \hat\rho_{\beta}\otimes\ket{+}\!\bra{+}$. We apply the controlled unitary evolution \begin{equation} \label{eq:hm}
\hat M_{\mu} = \hat U_{\phi}(T)e^{-\ii\mu \hat H_0}\otimes\ket{0}\!\bra{0} + e^{-\ii\mu \hat H_0}\hat U_{\phi}(T)\otimes\ket{1}\!\bra{1},
\end{equation} where $\hat U_{\phi}(T)$ is the unitary on the field generated by the Hamiltonian \eqref{eq:hamiltonian}, given by \begin{equation}
\label{eq:unitary}
    \hat U_{\phi}(T) = \mathcal{T} \exp\left(-\ii\lambda\!\int_{\mathbb{R}}\!\dd t\, \chi(t)\!\int_{\mathbb{R}^3} \!\!\!\dd^3\bm{x}\, F(\bm{x})\hat \phi (t,\bm{x})\right),
\end{equation}
where $\mathcal{T}$ represents time-ordering. 
Assuming that the coupling $\lambda$ is small enough, we can obtain an approximate expression for $\hat U_{\phi}(T)$ through a Dyson expansion: $
    \hat U_{\phi}(T) = \openone + \hat U^{(1)} + \hat U^{(2)} + \mathcal{O}(\lambda^3)$,
where in the interaction picture
\begin{equation}
\label{eq:pert}
    \hat U^{(1)} \!=\! -\ii\lambda\int_\mathbb{R}\!\! \dd t\, \hat H_I(t),\; \hat U^{(2)}\!=\! -\lambda ^2\!\! \int_{-\infty}^{\infty}\!\!\!\!\!\!\dd t\!\int_{-\infty}^{t}\!\!\!\!\!\!\dd t' \hat H_I(t)\hat H_I(t').
\end{equation}

% \nemma{Why is this unitary operator repeated? is \eqref{eq:hm} supposed to be the state of the ancillary?} \alvo{It shouldnt be there} \educ{It's fixed now. Those mistakes were introduced when removing the vacuum case}
The reduced state of the qubit at time $T$ can be written as $\hat \rho_{T} = \hat \rho_{T}^{(0)} + \hat \rho_{T}^{(1)} + \hat \rho_{T}^{(2)} + \mathcal O(\lambda^3)$, where $\hat \rho_{T}^{(i)}$ is proportional to $\lambda^i$  (The explicit expression can be seen in Appendix~\ref{app:kms})
%As shown in Appendix \ref{app:vacuum}, the reduced state can be written \st{nicely} as

 %\begin{align}
  %  &\hat \rho_{\mu} = \frac{1}{2}\bigg[\openone + \ket{-}\!\bra{+}\bigg(1 + \lambda^2\int\frac{\dd^3\bm{k}}{(2\pi)^3 2\omega_{\bm k}}|\widetilde{\chi}(\omega_{\bm k})|^2|\widetilde{F}(\bm{k})|^2 \nonumber \\
  %& \times \big(e^{-\ii\mu \omega_{\bm k}} - 1\big)\bigg) + 
   % \ket{+}\!\bra{-}\bigg(1 + \lambda^2\int\frac{\dd^3\bm{k}}{(2\pi)^3 2\omega_{\bm k}}|\widetilde{\chi}(\omega_{\bm k})|^2 \nonumber \\
   %&\times|\widetilde{F}(\bm{k})|^2 \big(e^{\ii\mu \omega_{\bm k}} - 1\big)\bigg)\bigg].
%\end{align}

 $\Tr[\hat\sigma_z\hat\rho_{\mu}]$ and $\Tr[\hat\sigma_y\hat\rho_{\mu}]$ give the real and imaginary parts, respectively, of the characteristic function \eqref{eq:characteristic}. Using the KMS two-point correlator  (see e.g., \cite{petar}), we can write the characteristic function for this process as
%We use that $\hat\sigma_z\ket{+} = \ket{-}, \hat\sigma_z\ket{-} = \ket{+}, \hat\sigma_y\ket{+} = -\ii\ket{-}$ and $\hat\sigma_y\ket{-} = \ii\ket{+}$. This yields, finally, that the characteristic function is
%\begin{align}
%\label{al:ch}
 %   \widetilde{P}(\mu) \coloneqq1 &+ \lambda^2\!\int\!\frac{\dd^3\bm{k}}{(2\pi)^3 2\omega_{\bm k}}|\widetilde{\chi}(\omega_{\bm k})|^2|\widetilde{F}(\bm{k})|^2\left(\cos(\mu \omega_{\bm k})\!-\! 1\right) \nonumber \\
  %  &+ \ii\lambda^2\!\int\!\frac{\dd^3\bm{k}}{(2\pi)^3 2\omega_{\bm k}}|\widetilde{\chi}(\omega_{\bm k})|^2|\widetilde{F}(\bm{k})|^2\sin(\mu \omega_{\bm k}).
%\end{align}
\begin{align}
\label{al:ch}
    \widetilde{P}(\mu) \coloneqq 1 + \lambda^2\int&\frac{\dd^3\bm{k}}{(2\pi)^32\omega_{\bm k}\left(e^{\beta \omega_{\bm k}} - 1\right)}|\widetilde{\chi}(\omega_{\bm k})|^2|\widetilde{F}(\bm{k})|^2 \nonumber \\
    &\times\left(e^{\beta \omega_{\bm k}} + 1\right)\left(\cos(\mu \omega_{\bm k}) - 1\right)  \\
   &\!\!\!\!\!\!\!\!\!\!\!\!\!\!\!\!\!\!\!\! + \ii\lambda^2\int\frac{\dd^3\bm{k}}{(2\pi)^32\omega_{\bm k}}|\widetilde{\chi}(\omega_{\bm k})|^2|\widetilde{F}(\bm{k})|^2\sin(\mu \omega_{\bm k}).\nonumber 
\end{align}

By taking the inverse Fourier transform of this characteristic function, the work probability distribution can be obtained. When the  smearing function is spherically symmetric and the field is massless, it is
%\begin{equation}
%\label{eq:p}
 %   P(W) = (1-p)\delta(W) + \frac{\lambda^2}{2\pi}|\widetilde{\chi}(W)|^2|\widetilde{F}(W)|^2W\Theta(W),
%\end{equation}
\begin{align}
    \label{al:pk}
    P(W) = (1-p)\delta(W) + \frac{\lambda^2}{2\pi}|\widetilde{\chi}(W)|^2|\widetilde{F}(W)|^2W \nonumber \\
    \times\left(\frac{e^{\beta W}}{e^{\beta W} - 1}\Theta(W) + \frac{1}{1-e^{-\beta W}}\Theta(-W)\right),
\end{align}
where $p \coloneqq \int_{W \neq 0}\dd W P(W)$ and $\Theta(W)$ is the Heaviside function. Note that the case of the vacuum state of the field can be obtained by taking the well-defined limit \mbox{$\beta\to\infty$} on Eq. \eqref{al:pk}.% To go from $\eqref{al:ch}$ to $\eqref{eq:p}$, we do a change of variables to spherical coordinates and then change the order the integration between $k$ and the Fourier variable $\mu$. Using the expressions for the inverse Fourier transform of the sine and the cosine, we get to the desired result.

\begin{figure}[h]
    \includegraphics[width = 0.4\textwidth]{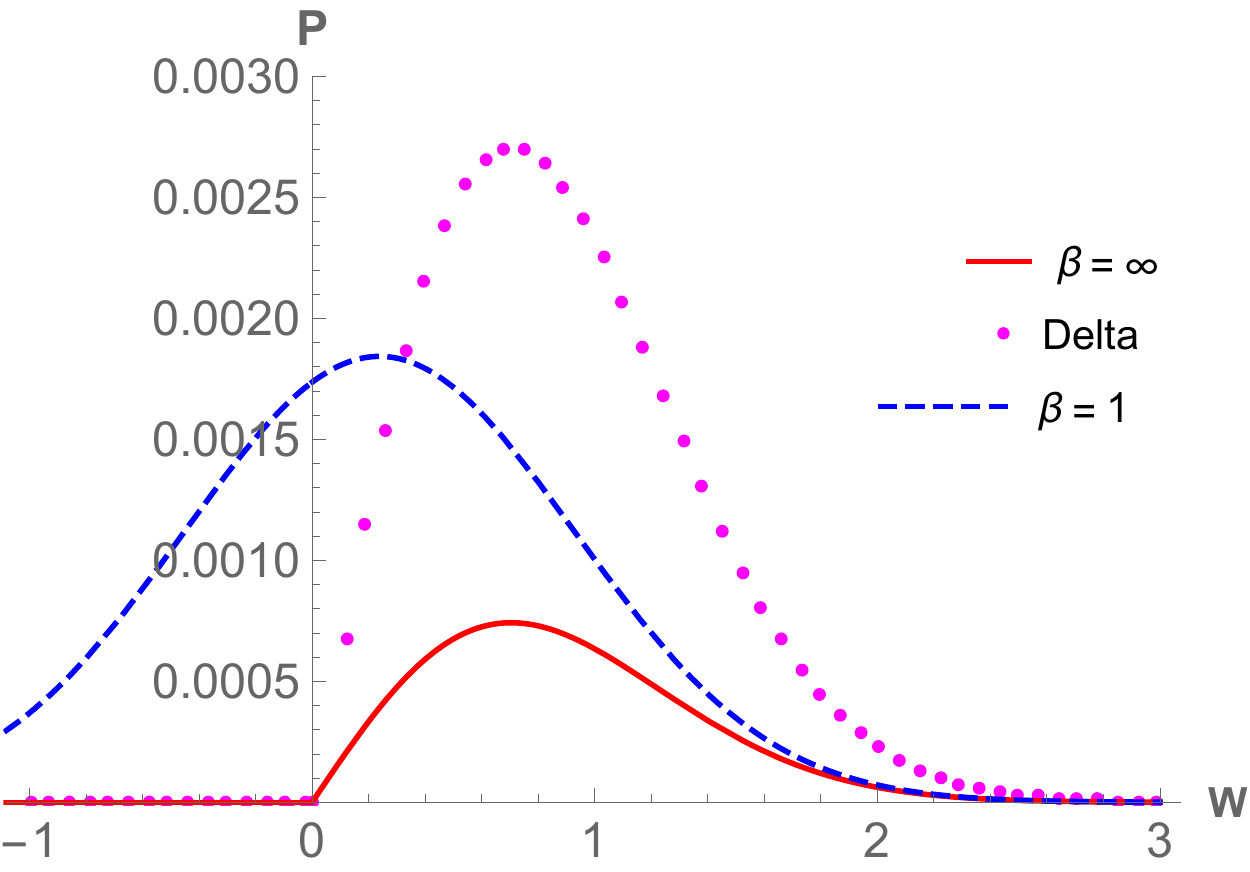}    \caption{Work distribution for three cases a) a localized unitary acting on the vacuum b) the same loicalized unitary acting on a KMS states of finite temperature and c) delta-coupling unitary acting on the vacuum. For a) and b) the switching function is of the form $\chi(t) = \exp[-(t - \frac{T}{2})^2(T^2/72)^{-1}]$, with $T = 1$ and $F(\bm x)$ is a normalized Gaussian distribution with $\sigma = 1$. Note that the length of the interval $[0,T]$ is 12 times the standard deviation of the switching function. For c) $F(\bm{x})$ is a normalized Gaussian distribution with $\sigma = 1$.}
    \label{fig:vacuum}
\end{figure}

In Fig. \ref{fig:vacuum}, we plot the work distribution for the unitary \eqref{eq:unitary} (omitting the deltas at the origin) acting on initial KMS states with $\beta=1$ and the vacuum state ($\beta\rightarrow \infty$),  for a particular choice of the switching and smearing functions. As  shown in Fig. \ref{fig:vacuum}, for the non-zero temperature states, there is a nonzero probability of the field doing work against the performer of the unitary, $W < 0$.  However, the probability of $W>0$ is larger than the probability of $W<0$, as granted by the second law.  For the vacuum case the performer of the unitary always has to work. As the duration of the process goes to infinity, the probability distribution gets concentrated around zero and the negative part of the distribution vanishes, as expected in the quasi-static limit. 

From \eqref{al:ch}, we can  now calculate the moments of $P(W)$ to gain some insight about the energy cost of applying a localized unitary to a quantum field. Since \mbox{$\widetilde{P}(\mu) =  \langle e^{\ii\mu W}\rangle$}, the $k$-th moment is\begin{equation}
\label{eq:moment}
    \langle W^k \rangle = \ii^{-k}\frac{\dd^k}{\dd^k\mu}\widetilde{P}(\mu)|_{\mu = 0}.
\end{equation}

 From \eqref{eq:moment} and \eqref{al:ch}, we obtain that the first and second moments of the work distribution for the vacuum are:
\begin{align}
    \langle W \rangle &= \lambda^2\int \frac{\dd^3\bm{k}}{(2\pi)^3 2}|\widetilde{\chi}(\omega_{\bm k})|^2|\widetilde{F}(\bm{k})|^2, \\
\langle    W^2 \rangle &= \lambda^2\int \frac{\dd^3\bm{k}}{(2\pi)^3 2}|\widetilde{\chi}(\omega_{\bm k})|^2|\widetilde{F}(\bm{k})|^2\omega_{\bm k},
\end{align}
obtaining $\sigma^2_W = \langle W^2\rangle - \langle W \rangle^2=
    \langle W^2 \rangle + \mathcal{O}(\lambda^4)$ for the work variance.
    
An interesting observation is that, for the vacuum, if we consider unitaries that are very localized in time and space, both $\widetilde{\chi}(\omega_{\bm{k}})$ and $\widetilde{F}(\bm{k})$  will be wide in frequency space, which means that the work variance will become larger than the expectation value, making the variance of the work  increasingly significant as the operation on the field becomes increasingly localized in both time and space.

For an arbitrary KMS state of inverse temperature $\beta$, the value for $\langle W \rangle$ coincides with that of the vacuum (and $\langle W \rangle\ge0$ as expected from the passivity of KMS states). In fact, since the imaginary part of the characteristic function does not depend on $\beta$, none of the odd-numbered moments will depend on temperature. For the variance, we have \begin{align}
\label{eq:var}
    % \sigma_{\beta}^2 = \frac{\lambda^2}{2}\!\!\int\!\!\frac{\dd^3\bm{k}}{(2\pi)^3\left(e^{\beta \omega_{\bm k}} - 1\right)}|\widetilde{\chi}(\omega_{\bm k})|^2|\widetilde{F}(\bm{k})|^2\left(e^{\beta \omega_{\bm k}}\!+\! 1\right)\omega_{\bm k},\\
    \sigma_{\beta}^2 = \frac{\lambda^2}{2}\!\!\int\!\!\frac{\dd^3\bm{k}}{(2\pi)^3}\frac{e^{\beta \omega_{\bm k}}\!+\! 1}{e^{\beta \omega_{\bm k}} - 1}|\widetilde{\chi}(\omega_{\bm k})|^2|\widetilde{F}(\bm{k})|^2\omega_{\bm k} + \mathcal{O}(\lambda^4),
\end{align}
 showing that it monotonically increases with temperature. 

We can also check that  Crooks' theorem \cite{crooks1999entropy} is satisfied for these localized unitaries. The theorem states that for a process in which the Hamiltonian evolves from $\hat H(0)=\hat H_1$ to $H(T)=\hat H_2$, together with its time-reversed process, we have that 
\begin{equation}
\label{eq:crooks}
\frac{P(W)}{P_{\text{rev}}(-W)}=e^{\beta W}\frac{Z_2}{Z_1},
\end{equation}
where $Z_1$, $Z_2$ are the partition functions of the thermal states of $\hat H(t_1)$ and $\hat H(t_2)$ and  the initial state must be thermal in both processes, with the corresponding Hamiltonian.

In our example, \mbox{$\widetilde P(\mu)=\widetilde{P}_{\text{rev}}(-\mu+\ii \beta)$} from equation \eqref{al:ch}, and since $\hat H(0)=\hat H(T)=\hat H_0$, $Z_2/Z_1=1$. Thus by taking the inverse Fourier transform we recover Eq. \eqref{eq:crooks}. Finally, the Jarzynski equality $\langle e^{-\beta W} \rangle = 1$, which is implied from the Crooks theorem, is satisfied. This can be seen just by evaluating the characteristic function at $\mu=\ii\beta$.

% ------------------------------------------ Delta coupling

\textit{\textbf{A non-perturbative example.-}} The examples in the previous section used small perturbations acting on thermal states only for calculational convenience. However, the work distribution we introduced is not limited to perturbative scenarios. Indeed, one of the main aims of fluctuation theorems is precisely to go beyond the regime of small perturbations by providing relations that hold for states arbitrarily far from equilibrium.

To illustrate this, we consider an intense unitary applied on the field very fast on a spatial distribution given by $F(\bm x)$. In this case, $\chi(t) = \delta(t)$, and the unitary in \eqref{eq:unitary} becomes to $\hat{U}_{\phi}(T) = \exp (-i\lambda\int d^3\textbf{x}F(\textbf{x})\hat{\phi}(\textbf{x}))$. Following the Ramsey scheme protocol, and using the non-perturbative techniques detailed in \cite{nonpert}, it is possible to obtain closed forms for  the characteristic function of the work distribution.
\begin{equation}
    \label{eq:deltac}
    \widetilde{P}(\mu) = \exp[\lambda^2\int \frac{\dd^3\textbf{k}}{(2\pi)^{3}2\omega_{\textbf{k}}}|\widetilde{F}(\textbf{k})|^2(e^{i\mu\omega_{\textbf{k}}} - 1)].
\end{equation}
The details of the calculation can be found in the Appendix~\ref{app:vacuum}.

 %By expanding in powers of $\lambda$  \eqref{eq:deltac} it is possible to check that at second order \eqref{eq:deltac} coincides with \eqref{al:ch} when $\beta = \infty$ and $\chi(t) = \delta(t)$.
Choosing a normalized Gaussian centered at zero as smearing, and changing to polar coordinates (since the smearing is spherically symmetrical), allows us to compute the characteristic function in a simple way (details in Appendix~\ref{app:vacuum}), yielding
\begin{align}
    \widetilde{P}(\mu) = e^{^{\frac{-\lambda^2}{8 \pi ^2\sigma^2}}}e^{ -\frac{\lambda^2e^{-\frac{\mu ^2}{4\sigma^2}} \left(2 e^{\frac{\mu ^2}{4\sigma^2}} \mu\sigma  \mathcal{D}\left(\frac{\mu }{2\sigma}\right)-2 e^{\frac{\mu^2}{4\sigma^2}}\sigma^2-\ii \sqrt{\pi } \mu \sigma \right)}{\left(4 \pi ^2\right) 4\sigma^4}},
\end{align}
where $\mathcal{D}(x)$ is the Dawson integral \cite{Abramowitz}, defined as $\mathcal{D}(x) \coloneqq \exp(-x^2)\int_0^x \exp(y^2)\dd y$. By taking the inverse Fourier transform numerically, we see how Crooks' theorem \eqref{eq:crooks} is also satisfied in this case, as $P(W) = 0$ for $W < 0$ (note that this is Crooks theorem for $\beta\rightarrow\infty$ when the forward and reverse processes are identical), as we show in Fig.~\ref{fig:vacuum}.

\textit{\textbf{Conclusion.-}}  The notion of work distributions for localized operations on quantum fields is challenging because a) energy eigenstates are not localized and b) projective measurements cannot be allowed in a relativistic quantum theory \cite{sorkin,Dowker,Dowker2}. The  TPM scheme employed in the literature \cite{Baeumer2018} is hence ill-defined in QFT, but we have shown that one can still make sense of it via the Ramsey scheme that was designed to measure TPM work distributions \cite{Dorner2013,Mazzola2014}. We  propose a well-defined work distribution in  QFT that, unlike \cite{bartolotta2018jarzynski}, does not require the existence of projective measurements and does not inherit any complications from the fact that energy eigenstates are non-local. We have shown that this work distribution satisfies both the Jarzynski equality and Crooks' theorem for  KMS states   for a general class of perturbative unitary actions arbitrarily localized in space and time. We also explicitly showed how Crooks theorem is satisfied for a general class of fast non-perturbative actions on the field vacuum generated by localized observables. These are limited cases. Showing that Crooks and Jarzynski are satisfied in the most general case is a non-trivial problem. It is known that for non-relativistic quantum systems, unitary operations acting on Gibbs thermal states implies satisfaction of these theorems \cite{campisicol,Albash13}, but showing whether this is true for all unitaries on all KMS states of fully relativistic field theories will require advanced tools from algebraic quantum field theory \cite{algebra,Fewster}. This is an interesting question that should be addressed elsewhere but is out of the scope of this letter.

%{\color{red} in both perturbative and non-perturbative regimes,} 

The proposed work distribution also suggests experiments where it can be measured. A potential setup would be a quantum field in an superconducting transmission line to which we couple superconducting qubits. The control in time that is required for an experiment implementing the example that we present in the manuscript can be achieved with the switchable coupling that has been experimentally realized in \cite{experiment}. The fact that the Ramsey scheme can be implemented in superconducting circuits was shown in, e.g, \cite{circuit}, and the fact that a fully-relativistic QFT setup is implementable in superconducting circuits in those regimes can be seen, e.g., in \cite{pastfuture,entanglementacceleration,fermion}.

An interesting observation is that the work distribution that we define can be used to compute ratios of partition functions of field theories. Indeed we can invert the relationship \eqref{eq:crooks} and write 
\begin{equation}
\frac{Z_2}{Z_1}=e^{-\beta W}\frac{P(W)}{P_{\text{rev}}(-W)}.
\end{equation}
This can in fact be more simply obtained from Jarzynski's equality
\begin{equation}
    \frac{Z_2}{Z_1}=\left\langle e^{-\beta W}\right\rangle.
\end{equation}
 This potentially provides a new way to compute these ratios,  analytically, numerically or even experimentally by measuring the work distribution through a Ramsey scheme. These ratios are remarkably difficult to calculate in QFT through path integrals, which makes new methods to access it a research avenue that merits exploration. The idea of calculating the ratio of partition functions from a non-equilibrium process has been used  in very different contexts (see e.g. \cite{liphardt2002equilibrium,park2003free,collin2005verification}).

 With our framework, we have been able to obtain expressions for the work fluctuations associated to a process generated by a local Hamiltonian on a scalar field. We observe that the work fluctuations increase with temperature, and that they dominate the average work cost as the process becomes increasingly localized in both time and space. Also, we find that for KMS states of finite temperature, there is a non-zero probability of the field doing work when the process is of finite duration. It should be interesting to see how the work distribution relates to the variation of internal energy in the field in adiabatic and non-adiabatic processes. The internal energy of the field is given by the re-normalized stress-energy density, and exploring the connection between the stress-energy density deposited (or extracted) from the field and the work distributions of the processes where the energy is deposited can shed some light into the thermodynamics of local processes in quantum field theory, a particularly relevant aspect of phenomena ranging from entanglement harvesting \cite{entanglealex,entangbell,entvac, nonlocal}, quantum energy teleportation \cite{qet}, or the Unruh and Hawking effects \cite{waiting}.

%{\color{magenta} Our findings suggest that the proposed work distribution together with either the Crooks theorem or the Jarzynski equality can be used as a new method for evaluating ratios of partition functions between two field theories connected by some one-parametric variation of their Hamiltonians. This opens the door to a new method of computing this ratio alternative to the usual path-integral formalism and may be worth exploring in future works.}\alv{if we need to cut words, we can merge this with the last paragraph of the previous section}

\acknowledgments
E. M-M. is supported by his Ontario Early Researcher Award and the NSERC Discovery program. Research  at  Perimeter  Institute  is  supported  by  the
Government of Canada through the Department of Innovation, Science and Economic Development and by the Province of Ontario through the Ministry of Research, Innovation and Science. A. O. acknowledges the support of Fundacio Privada Cellex, through a Mobility Research Award.

% ------------------------------------------ APPENDICES       

\appendix

\begin{widetext}
\section{Details of the calculation of the state of the qubit}\label{app:vacuum}

\subsection{Field in the vacuum state}

We now proceed to calculate the different terms of the perturbative expansion in (9) in the main text. Clearly, $\hat\rho_{T}^{(0)} = \Tr_{\phi}\left(\hat\rho_{0}\right) = \ket{+}\!\bra{+}$. It is easy to see that the first order term $\hat\rho_T^{(1)}$ will vanish. When taking the trace over the field, all the free evolution terms will end up multiplying the vacuum state, either at their left or at their right, so they will disappear, leaving $\bra{\Omega}\hat U^{(1)}\ket{\Omega}$ or $\bra{\Omega}\hat U^{\dagger (1)}\ket{\Omega}$. This is zero since \begin{equation}
    \label{eq:u1}
    \hat U^{(1)} = -\ii\lambda\int_{-\infty}^{\infty}\dd t\int \dd^3\bm{x} \chi(t)F(\bm{x})\hat\phi(t,\bm{x}),
\end{equation}
and $\bra{\Omega}\hat\phi(t,\bm{x})\ket{\Omega} = 0$ $\forall$ $t$, $\bm{x}$.

$\hat\rho_T^{(2)}$ is the sum of two contributions, one involving products with $\hat U^{(1)}$ and $\hat U^{\dagger (1)}$, and the other with $\hat U^{(2)}$. Let us focus on the first family of terms.\\

As an example, we explicitly calculate the coefficient associated to the component $\frac{\ket{0}\!\bra{0}}{2}$ of the density matrix of the qubit. 
\begin{align} \nonumber
\Tr\left(\hat U^{(1)}e^{-\ii\mu \hat H_0}\ket{\Omega}\!\bra{\Omega}e^{\ii\mu \hat H_0}\hat U^{\dagger (1)}\right) &= \Tr\left(\hat U^{(1)}\ket{\Omega}\!\bra{\Omega}\hat U^{\dagger (1)}\right)\\ \nonumber
   &=  \Tr\left(\lambda^2\int_{-\infty}^{\infty}\dd t\chi(t)\int_{-\infty}^{\infty}\dd t'\chi(t')\int \dd^3\bm{x} F(\bm{x})\int \dd^3\bm{x}' F(\bm{x}')\hat\phi(t,\bm{x})\ket{\Omega}\!\bra{\Omega}\hat\phi(t',\bm{x}')\right)\\ \nonumber
    &= \Tr\left(\lambda^2\int\frac{\dd^3\bm{k} \dd^3\bm{k}'}{(2\pi)^3\sqrt{2\omega_{\bm k}}\sqrt{2\omega_{\bm{k}'}}}\int_{-\infty}^{\infty}\dd t\chi(t)e^{i\omega_{\bm k}t}\int \dd^3\bm{x} F(\bm{x})e^{-\ii\bm{k}\bm{x}} \right. \\ \nonumber
    &\qquad\qquad\qquad\textbf{}\times\left.\int_{-\infty}^{\infty}\dd t'\chi(t')e^{-\ii \omega_{\bm{k}'}t'}\int \dd^3\bm{x}' F(\bm{x}')e^{\ii\bm{k}'\cdot\bm{x}'}|\bm{k}\rangle \langle\bm{k'}|\right)\\ \nonumber
    &= \lambda^2\Tr\left(\int\frac{\dd^3\bm{k}\dd^3\bm{k}'}{(2\pi)^3\sqrt{2\omega_{\bm k}}\sqrt{2\omega_{\bm{k}'}}}\widetilde{\chi}(\omega_{\bm k})\widetilde{\chi}(-\omega_{\bm{k}'})\widetilde{F}(-\bm{k})\widetilde{F}(-\bm{k'})|\bm{k}\rangle\langle\bm{k'}|\right) \\
    \label{eq:o1}
    &=  \lambda^2\int\frac{\dd^3k}{(2\pi)^32\omega_{\bm k}}|\widetilde{\chi}(\omega_{\bm k})|^2|\widetilde{F}(\bm{k})|^2,
\end{align}
where in the last step we have used that $FT[f](x) = (FT[f](-x))^*$, for a real function $f$. We are assuming that both the switching and the smearing are real functions. The calculation for the $\frac{\ket{1}\!\bra{0}}{2}$ coefficient is analogous, the only difference being the presence of a factor $e^{-i\mu  \hat H_0}$ multiplying the ket vectors $| \bm{k}\rangle$. Since $e^{-\ii\mu \hat H_0}| \bm{k}\rangle = e^{-\ii\mu \omega_{\bm k}}|\bm{k}\rangle$, we obtain that \begin{equation}
    \Tr\left(e^{-\ii\mu \hat H_0}\hat U^{(1)}\ket{\Omega}\!\bra{\Omega}e^{\ii\mu \hat H_0}\hat U^{\dagger(1)}\right) = \lambda^2\int\frac{\dd^3\bm{k}}{(2\pi)^32\omega_{\bm k}}|\widetilde{\chi}(\omega_{\bm k})|^2|\widetilde{F}(\bm{k})|^2e^{-\ii\mu \omega_{\bm k}}.
\end{equation} 
The rest of the components are the Hermitian conjugates of these.

We now calculate the remaining terms. That is, the terms that involve products with $\hat U^{(2)}$. Let us start by obtaining $\Tr_{\phi}\left(\hat U^{(2)}e^{-\ii\mu \hat H_0}\ket{\Omega}\!\bra{\Omega}e^{\ii\mu \hat H_0}\otimes\frac{\ket{0}\bra{0}}{2} + \text{H.c} \right) = \bra{\Omega}\hat U^{(2)}\ket{\Omega}\frac{\ket{0}\!\bra{0}}{2} + H.c$. This is simply
\begin{align}
    &-\lambda^2\int_{-\infty}^{\infty}\dd t\int_{-\infty}^{t}\dd t'\int \dd^3 \bm{x}\int \dd^3 \bm{x}' \chi(t)\chi(t')F(\bm{x})F(\bm{x}')\bra{\Omega}\hat\phi(t,\bm{x})\hat\phi(t',\bm{x}')\ket{\Omega}\frac{\ket{0}\!\bra{0}}{2} + H.c  \\ \nonumber
    &\hspace{3cm}=-\lambda^2\int_{-\infty}^{\infty}\dd t\int_{-\infty}^{t}\dd t'\int \dd^3 \bm{x}\int \dd^3 \bm{x}' \chi(t)\chi(t')F(\bm{x})F(\bm{x}')2\Re[\mathcal{W}(t,\bm{x},t',\bm{x}')]\frac{\ket{0}\!\bra{0}}{2}, 
\end{align}
where $\mathcal{W}(t,\bm{x},t',\bm{x}')=\bra{\Omega}\hat\phi(t,\bm{x})\hat \phi(t',\bm{x}')\ket{\Omega}$ is the Wightman function. The same is obtained for the other cases. This is because all the $e^{\pm i\mu \hat H_0}$ end up multiplying the vacuum state when taking the trace, so they disappear leaving simply $\bra{\Omega}\hat U^{(2)}\ket{\Omega} + H.c$. Therefore, the contribution of these terms to the reduced state of the qubit is $A\ket{+}\bra{+}$, where \begin{equation}
\label{eq:A}
    A =  -\lambda^2\int_{-\infty}^{\infty}\dd t\int_{-\infty}^{t}\dd t'\int \dd^3 \bm{x}\int \dd^3 \bm{x}' \chi(t)\chi(t')F(\bm{x})F(\bm{x}')2\Re[\mathcal{W}(t,\bm{x},t',\bm{x}')].
\end{equation}

Adding everything and noting that \eqref{eq:o1} is equal to \eqref{eq:A} we obtain, after applying the second Hadamard on the qubit, that the reduced state can be written as
 \begin{align}
    \hat\rho_{\mu} = \frac{1}{2}\left(\mathbb{I} + \ket{-}\bra{+}\left(1 + \lambda^2\int\frac{\dd^3\bm{k}}{(2\pi)^3 2\omega_{\bm k}}|\widetilde{\chi}(\omega_{\bm k})|^2|\widetilde{F}(\bm{k})|^2\left(e^{-\ii\mu \omega_{\bm k}} - 1\right)\right) \right.  \\
    \left.+
    \ket{+}\bra{-}\left(1 + \lambda^2\int\frac{\dd^3\bm{k}}{(2\pi)^3 2\omega_{\bm k}}|\widetilde{\chi}(\omega_{\bm k})|^2|\widetilde{F}(\bm{k})|^2\left(e^{\ii\mu \omega_{\bm k}} - 1\right)\right)\right).
\end{align}

\subsection{Field in a finite-temperature KMS state}
    \label{app:kms}
    
Some properties that we use throughout these calculation are: \begin{align}
    \label{eq:ort}
    \Tr(\hat\phi \hat\rho_{\beta}) &= 0,\\
\label{eq:com}
    \left[\hat\rho_{\beta}, e^{-it\hat H_0}\right] &= 0.
\end{align}
As before, we calculate the reduced state of the qubit with a Dyson expansion. The first order term is again zero. As an example of why this is the case, we calculate
\begin{align}
    \Tr\left(e^{-\ii\mu \hat H_0}\hat U^{(1)}\hat\rho_{\beta} e^{\ii\mu \hat H_0}\right) + \Tr\left(e^{-\ii\mu \hat H_0}\hat\rho_{\beta}e^{\ii\mu \hat H_0}\hat U^{\dagger (1)}\right)
    = \Tr\left(\hat U^{(1)}\hat\rho_{\beta}\right) + Tr\left(\hat\rho_{\beta} \hat U^{\dagger (1)}\right),
\end{align}
where we have used the cyclic property of the trace and \eqref{eq:com}. Finally, using the linearity of the trace, the expression for $\hat U^{(1)}$ and \eqref{eq:ort}, we obtain that both terms are zero. A similar procedure can be used to check that all the other contributions to the first order correction are zero.

We now calculate the second order terms, starting by the ones that only involve products of $\hat U^{(1)}$ and $\hat U^{\dagger(1)}$. We derive here only the coefficient of $\frac{\ket{0}\bra{1}}{2}$. The other cases follow analogously.
 \begin{align}
\label{al:2}
    \Tr\left(\hat U^{(1)}e^{-\ii\mu \hat H_0}\hat\rho_{\beta} \hat U^{\dagger (1)}e^{\ii\mu \hat H_0}\right) = \Tr\left(e^{\ii\mu \hat H_0}\hat U^{(1)}e^{-\ii\mu \hat H_0}\hat\rho_{\beta} \hat U^{\dagger (1)}\right).
\end{align}
We have that \begin{align}
    e^{\ii\mu \hat H_0}\hat U^{(1)}e^{-\ii\mu \hat H_0} &= -\ii\lambda\int_{-\infty}^{\infty}\dd t\int \dd^3 \bm{x}\chi(t)F(\bm{x})e^{\ii\mu \hat H_0}\hat\phi(\bm{x},t)e^{-\ii\mu  \hat H_0} \\ \nonumber
    &= -\ii\lambda\int_{-\infty}^{\infty}\dd t\int \dd^3 \bm{x}\chi(t)F(\bm{x})\hat\phi(\bm{x},t + \mu).
\end{align}
So \eqref{al:2} equals \begin{align}
\label{al:ub1}
    \lambda^2\int \dd t\int \dd t'\int \dd^3 \bm{x}\int \dd^3 \bm{x}'\chi(t)\chi(t')F(\bm{x})F(\bm{x}')\Tr\left(\hat\phi(\bm{x},t + \mu)\hat\rho_{\beta}\hat\phi(\bm{x}', t' + \mu)\right) \\ \nonumber
    = \lambda^2\int \dd t\int \dd t'\int \dd^3 \bm{x}\int \dd^3 \bm{x}'\chi(t)\chi(t')F(\bm{x})F(\bm{x}')\mathcal{W}_{\beta}(\bm{x}',t',\bm{x},t + \mu).
\end{align}
The other terms have the same structure, with the only change being in the thermal Wightman function, which is $\mathcal{W}(x',t',x,t)$ for the diagonal terms, and $\mathcal{W}(x',t',x,t - \mu)$ for the $\frac{\ket{1}\!\bra{0}}{2}$ term.

Let us obtain now the second order terms coming from products with $\hat U^{(2)}$ and $\hat U^{\dagger (2)}$. It is easy to see, using \eqref{eq:com} that all the terms are equal to \begin{equation}
\label{eq:ub2}
    \Tr\left(\hat U^{(2)}\hat\rho_{\beta}\right) + \Tr\left(\hat\rho_{\beta} \hat U^{\dagger (2)}\right) = 2\Re\Tr\left(\hat U^{(2)}\hat\rho_{\beta}\right).
\end{equation}
This finishes the proof. Since the Dyson expansion preserves the trace of the density matrix, Eq. \eqref{al:ub1} has to be equal to Eq. \eqref{eq:ub2}, so as to cancel the diagonals added by the perturbation terms. This is useful because it gives a much more compact expression for the reduced state of the qubit. Therefore, at the end of the Ramsey scheme, the density matrix of the qubit is: \begin{align}
    &\frac{1}{2}\left(\mathbb{I} + \ket{+}\bra{-}\left(1 + \lambda^2\int \dd t\int \dd t'\int \dd^3 \bm{x}\int \dd^3 \bm{x}'\chi(t)\chi(t')F(\bm{x})F(\bm{x}')\left(\mathcal{W}_{\beta}(\bm{x}',t',\bm{x},t + \mu) - \mathcal{W}_{\beta}(\bm{x}',t',\bm{x},t)\right)\right)\right. \\ \nonumber
    &\quad+ \ket{-}\bra{+}\left.\left(1 + \lambda^2\left(\int \dd t\int \dd t'\int \dd^3 \bm{x}\int \dd^3 \bm{x}'\chi(t)\chi(t')F(\bm{x})F(\bm{x}')\left(\mathcal{W}_{\beta}(\bm{x}',t',\bm{x},t - \mu) - \mathcal{W}_{\beta}(\bm{x}',t',\bm{x},t)\right)\right)\right)\right).
\end{align}  

Using the expression of the Wightman function for a thermal state of inverse temperature $\beta$ \cite{petar}
\begin{equation}
    \mathcal{W}_{\beta}(\bm{x}',t',\bm{x},t) = \int \frac{\dd^3\bm{k}}{(2\pi)^3 2\omega_{\bm k}\left(e^{\beta \omega_{\bm k}} - 1\right)}\left(e^{\beta \omega_{\bm k}}e^{\ii k\left(\bm x - \bm x'\right)} + e^{\ii k\left(\bm x' - \bm x\right)}\right),
\end{equation}
we can calculate the characteristic function of the work distribution \begin{align}
    \widetilde{P}(\mu) = 1 + \lambda^2\int&\frac{d^3\bm k}{(2\pi)^32\omega_{\bm k}\left(e^{\beta \omega_{\bm k}} - 1\right)}|\widetilde{\chi}(\omega_{\bm k})|^2|\widetilde{F}(\bm{k})|^2\left(e^{\beta \omega_{\bm k}} + 1\right)\left(\cos(\mu \omega_{\bm k}) - 1\right) \\ \nonumber
    &+ \ii\lambda^2\int\frac{d^3\bm k}{(2\pi)^32\omega_{\bm k}}|\widetilde{\chi}(\omega_{\bm k})|^2|\widetilde{F}(\bm{k})|^2\sin(\mu \omega_{\bm k}).
\end{align}

\section{Delta-coupling to the vacuum}
    \label{app:delta}
    
    Let us start by defining a general coherent state of the field $\ket{\alpha(\textbf{k})}$: \begin{equation}
        \label{eq:coherent}
        \ket{\alpha(\textbf{k})} = \hat{D}_{\alpha_{(\textbf{k})}}\ket{\Omega} = \exp(\int d^3{\textbf{k}}[\alpha(\textbf{k})\hat{a}_{\textbf{k}}^{\dagger} - \alpha^{*}(\textbf{k})\hat{a}_{\textbf{k}}])\ket{\Omega}.
    \end{equation}
    When the interaction Hamiltonian couples to the field through a delta switching function, the resulting unitary is \begin{equation}
    \hat{U}_{\phi}(T) = \exp (-i\lambda\int d^3\textbf{x}F(\textbf{x})\hat{\phi}(\textbf{x})) = \hat{D}_{\alpha_{(\textbf{k})}},
    \end{equation}
 with $\alpha_{(\textbf{k})} = \frac{-i\lambda \widetilde{F}^{*}(\textbf{k})}{(2\pi)^{3/2}\sqrt{2\omega_{\textbf{k}}}}$. This can be seen writing the field operator in its mode decomposition. We now calculate the reduced state of the qubit before applying the last Hadamard gate of the Ramsey scheme. The diagonal elements of the density matrix of the qubit are both $\frac{1}{2}$. The off-diagonal terms are $\frac{1}{2}\bra{\Omega}\hat{U}^{\dagger}_{\phi}(T)e^{-i\mu\hat{H}_0}\hat{U}_{\phi}(T)\ket{\Omega}$ and the Hermitian conjugate. 
 
 Using that $e^{-i\mu\hat{H}_0}\ket{\Omega} = \ket{\Omega}$, we can rewrite the previous expression as \begin{equation}
    \label{eq:inner}
     \bra{\Omega}\hat{U}^{\dagger}_{\phi}(T)e^{-i\mu\hat{H}_0}\hat{U}_{\phi}(T)e^{i\mu\hat{H}_0}\ket{\Omega} = \braket{\alpha(\textbf{k})|\beta(\textbf{k})},
 \end{equation}
 with $\beta(\textbf{k}) = \frac{-i\lambda\widetilde{F}^{*}\textbf{k}e^{-i\omega_{\textbf{k}}\mu}}{(2\pi)^{3/2}\sqrt{2\omega_{\textbf{k}}}}$, as can be seen by using that $e^{-i\mu \hat{H}_0}\hat{a}_{\bm{k}}e^{i\mu \hat{H}_0} = e^{i\mu\omega_{\bm{k}}}\hat{a}_{\bm{k}}$, and $e^{-i\mu \hat{H}_0}\hat{a}^{\dagger}_{\bm{k}}e^{i\mu \hat{H}_0} = e^{-i\mu\omega_{\bm{k}}}\hat{a}^{\dagger}_{\bm{k}}$ . Using the expression for the inner product of two coherent states in Appendix A of \cite{nonpert}, \eqref{eq:inner} can be simplified to \begin{equation}
     \braket{\alpha(\textbf{k})|\beta(\textbf{k})} = \exp[\lambda^2\int \frac{d^3\textbf{k}}{(2\pi)^{3}2\omega_{\textbf{k}}}|\widetilde{F}(\textbf{k})|^2(e^{-i\mu\omega_{\textbf{k}}} - 1)].
 \end{equation}
 
Applying the second Hadamard to the qubit and taking the Z and Y components of the Bloch vector of the qubit finally yields 
\begin{equation}
    \widetilde{P}(\mu) = \exp[\lambda^2\int \frac{\dd^3\textbf{k}}{(2\pi)^{3}2\omega_{\textbf{k}}}|\widetilde{F}(\textbf{k})|^2(e^{i\mu\omega_{\textbf{k}}} - 1)].
\end{equation}

Choosing a normalized spherical Gaussian centered at zero as smearing $F(r) = \frac{e^{-\frac{r^2}{2\sigma^2}}}{\sqrt{2\pi\sigma^2}}$, and changing to polar coordinates (since the smearing is spherically symmetrical), yields for the characteristic function of a massless scalar field ($\omega_{\bm{k}} = |\bm{k}|)$ \begin{equation}
    \widetilde{P}(\mu) = \exp[\frac{\lambda^2}{4\pi^2}\int_0^{\infty}\dd |\bm{k}||\widetilde{F}(|\bm{k}|)|^2|\bm{k}|(e^{i\mu|\bm{k}|} - 1)].
\end{equation} 
Using that \begin{equation}
    \int_0^{\infty}\dd |\bm{k}||\widetilde{F}(|\bm{k}|)|^2|\bm{k}| = \frac{1}{2\sigma^2},
\end{equation}
and \begin{equation}
    \int_0^{\infty} \dd |\bm{k}||\widetilde{F}(|\bm{k})|^2|\bm{k}|e^{i\mu|\bm{k}} = \exp \left(\frac{e^{-\frac{\mu ^2}{4\sigma^2}} \left(2 e^{\frac{\mu ^2}{4\sigma^2}} \mu\sigma  \mathcal{D}\left(\frac{\mu }{2\sigma}\right)-2 e^{\frac{\mu^2}{4\sigma^2}}\sigma^2-i \sqrt{\pi } \mu \sigma \right)}{4\sigma^4}\right),
\end{equation}
yields for the characteristic function
\begin{equation}
    \widetilde{P}(\mu) = \exp \left(-\lambda^2\frac{e^{-\frac{\mu ^2}{4\sigma^2}} \left(2 e^{\frac{\mu ^2}{4\sigma^2}} \mu\sigma  \mathcal{D}\left(\frac{\mu }{2\sigma}\right)-2 e^{\frac{\mu^2}{4\sigma^2}}\sigma^2-i \sqrt{\pi } \mu \sigma \right)}{\left(4 \pi ^2\right) 4\sigma^4}-\frac{\lambda^2}{8 \pi ^2\sigma^2}\right),
\end{equation}
 where $\mathcal{D}(x)$ is the Dawson Function, defined as $\mathcal{D}(x) = e^{-x^2}\int_0^x \dd y e^{y^2}$.
 \end{widetext}

\bibliography{bibliography}

\begin{thebibliography}{43}%
\makeatletter
\providecommand \@ifxundefined [1]{%
 \@ifx{#1\undefined}
}%
\providecommand \@ifnum [1]{%
 \ifnum #1\expandafter \@firstoftwo
 \else \expandafter \@secondoftwo
 \fi
}%
\providecommand \@ifx [1]{%
 \ifx #1\expandafter \@firstoftwo
 \else \expandafter \@secondoftwo
 \fi
}%
\providecommand \natexlab [1]{#1}%
\providecommand \enquote  [1]{``#1''}%
\providecommand \bibnamefont  [1]{#1}%
\providecommand \bibfnamefont [1]{#1}%
\providecommand \citenamefont [1]{#1}%
\providecommand \href@noop [0]{\@secondoftwo}%
\providecommand \href [0]{\begingroup \@sanitize@url \@href}%
\providecommand \@href[1]{\@@startlink{#1}\@@href}%
\providecommand \@@href[1]{\endgroup#1\@@endlink}%
\providecommand \@sanitize@url [0]{\catcode `\\12\catcode `\$12\catcode
  `\&12\catcode `\#12\catcode `\^12\catcode `\_12\catcode `\%12\relax}%
\providecommand \@@startlink[1]{}%
\providecommand \@@endlink[0]{}%
\providecommand \url  [0]{\begingroup\@sanitize@url \@url }%
\providecommand \@url [1]{\endgroup\@href {#1}{\urlprefix }}%
\providecommand \urlprefix  [0]{URL }%
\providecommand \Eprint [0]{\href }%
\providecommand \doibase [0]{http://dx.doi.org/}%
\providecommand \selectlanguage [0]{\@gobble}%
\providecommand \bibinfo  [0]{\@secondoftwo}%
\providecommand \bibfield  [0]{\@secondoftwo}%
\providecommand \translation [1]{[#1]}%
\providecommand \BibitemOpen [0]{}%
\providecommand \bibitemStop [0]{}%
\providecommand \bibitemNoStop [0]{.\EOS\space}%
\providecommand \EOS [0]{\spacefactor3000\relax}%
\providecommand \BibitemShut  [1]{\csname bibitem#1\endcsname}%
\let\auto@bib@innerbib\@empty
%</preamble>
\bibitem [{\citenamefont {Campisi}\ \emph {et~al.}(2011)\citenamefont
  {Campisi}, \citenamefont {H\"anggi},\ and\ \citenamefont
  {Talkner}}]{campisicol}%
  \BibitemOpen
  \bibfield  {author} {\bibinfo {author} {\bibfnamefont {M.}~\bibnamefont
  {Campisi}}, \bibinfo {author} {\bibfnamefont {P.}~\bibnamefont {H\"anggi}}, \
  and\ \bibinfo {author} {\bibfnamefont {P.}~\bibnamefont {Talkner}},\ }\href
  {\doibase 10.1103/RevModPhys.83.771} {\bibfield  {journal} {\bibinfo
  {journal} {Rev. Mod. Phys.}\ }\textbf {\bibinfo {volume} {83}},\ \bibinfo
  {pages} {771} (\bibinfo {year} {2011})}\BibitemShut {NoStop}%
\bibitem [{\citenamefont {Esposito}\ \emph
  {et~al.}(2009{\natexlab{a}})\citenamefont {Esposito}, \citenamefont
  {Harbola},\ and\ \citenamefont {Mukamel}}]{esposito}%
  \BibitemOpen
  \bibfield  {author} {\bibinfo {author} {\bibfnamefont {M.}~\bibnamefont
  {Esposito}}, \bibinfo {author} {\bibfnamefont {U.}~\bibnamefont {Harbola}}, \
  and\ \bibinfo {author} {\bibfnamefont {S.}~\bibnamefont {Mukamel}},\ }\href
  {\doibase 10.1103/RevModPhys.81.1665} {\bibfield  {journal} {\bibinfo
  {journal} {Rev. Mod. Phys.}\ }\textbf {\bibinfo {volume} {81}},\ \bibinfo
  {pages} {1665} (\bibinfo {year} {2009}{\natexlab{a}})}\BibitemShut {NoStop}%
\bibitem [{\citenamefont {Jarzynski}(2011)}]{jarz2}%
  \BibitemOpen
  \bibfield  {author} {\bibinfo {author} {\bibfnamefont {C.}~\bibnamefont
  {Jarzynski}},\ }\href {\doibase 10.1146/annurev-conmatphys-062910-140506}
  {\bibfield  {journal} {\bibinfo  {journal} {Annu. Rev. Condens. Matter Phys}\
  }\textbf {\bibinfo {volume} {2}},\ \bibinfo {pages} {329} (\bibinfo {year}
  {2011})}\BibitemShut {NoStop}%
\bibitem [{\citenamefont {Talkner}\ \emph {et~al.}(2007)\citenamefont
  {Talkner}, \citenamefont {Lutz},\ and\ \citenamefont {H\"anggi}}]{talkner}%
  \BibitemOpen
  \bibfield  {author} {\bibinfo {author} {\bibfnamefont {P.}~\bibnamefont
  {Talkner}}, \bibinfo {author} {\bibfnamefont {E.}~\bibnamefont {Lutz}}, \
  and\ \bibinfo {author} {\bibfnamefont {P.}~\bibnamefont {H\"anggi}},\ }\href
  {\doibase 10.1103/PhysRevE.75.050102} {\bibfield  {journal} {\bibinfo
  {journal} {Phys. Rev. E}\ }\textbf {\bibinfo {volume} {75}},\ \bibinfo
  {pages} {050102} (\bibinfo {year} {2007})}\BibitemShut {NoStop}%
\bibitem [{\citenamefont {{B{\"a}umer}}\ \emph {et~al.}(2018)\citenamefont
  {{B{\"a}umer}}, \citenamefont {{Lostaglio}}, \citenamefont
  {{Perarnau-Llobet}},\ and\ \citenamefont {{Sampaio}}}]{Baeumer2018}%
  \BibitemOpen
  \bibfield  {author} {\bibinfo {author} {\bibfnamefont {E.}~\bibnamefont
  {{B{\"a}umer}}}, \bibinfo {author} {\bibfnamefont {M.}~\bibnamefont
  {{Lostaglio}}}, \bibinfo {author} {\bibfnamefont {M.}~\bibnamefont
  {{Perarnau-Llobet}}}, \ and\ \bibinfo {author} {\bibfnamefont
  {R.}~\bibnamefont {{Sampaio}}},\ }\href@noop {} {\bibfield  {journal}
  {\bibinfo  {journal} {arXiv:1805.10096}\ } (\bibinfo {year}
  {2018})}\BibitemShut {NoStop}%
\bibitem [{\citenamefont {Tasaki}(2000)}]{tasaki2000jarzynski}%
  \BibitemOpen
  \bibfield  {author} {\bibinfo {author} {\bibfnamefont {H.}~\bibnamefont
  {Tasaki}},\ }\href@noop {} {\bibfield  {journal} {\bibinfo  {journal} {arXiv
  preprint cond-mat/0009244}\ } (\bibinfo {year} {2000})}\BibitemShut {NoStop}%
\bibitem [{\citenamefont {Kurchan}(2001)}]{referee}%
  \BibitemOpen
  \bibfield  {author} {\bibinfo {author} {\bibfnamefont {J.}~\bibnamefont
  {Kurchan}},\ }\href@noop {} {\bibfield  {journal} {\bibinfo  {journal} {arXiv
  preprint cond-mat/0007360}\ } (\bibinfo {year} {2001})}\BibitemShut {NoStop}%
\bibitem [{\citenamefont {Esposito}\ \emph
  {et~al.}(2009{\natexlab{b}})\citenamefont {Esposito}, \citenamefont
  {Harbola},\ and\ \citenamefont {Mukamel}}]{espositofluct}%
  \BibitemOpen
  \bibfield  {author} {\bibinfo {author} {\bibfnamefont {M.}~\bibnamefont
  {Esposito}}, \bibinfo {author} {\bibfnamefont {U.}~\bibnamefont {Harbola}}, \
  and\ \bibinfo {author} {\bibfnamefont {S.}~\bibnamefont {Mukamel}},\ }\href
  {\doibase 10.1103/RevModPhys.81.1665} {\bibfield  {journal} {\bibinfo
  {journal} {Rev. Mod. Phys.}\ }\textbf {\bibinfo {volume} {81}},\ \bibinfo
  {pages} {1665} (\bibinfo {year} {2009}{\natexlab{b}})}\BibitemShut {NoStop}%
\bibitem [{\citenamefont {Redhead}(1995)}]{Redhead1995}%
  \BibitemOpen
  \bibfield  {author} {\bibinfo {author} {\bibfnamefont {M.}~\bibnamefont
  {Redhead}},\ }\href {\doibase 10.1007/BF02054660} {\bibfield  {journal}
  {\bibinfo  {journal} {Found. Phys.}\ }\textbf {\bibinfo {volume} {25}},\
  \bibinfo {pages} {123} (\bibinfo {year} {1995})}\BibitemShut {NoStop}%
\bibitem [{\citenamefont {Sorkin}(1993)}]{sorkin}%
  \BibitemOpen
  \bibfield  {author} {\bibinfo {author} {\bibfnamefont {R.~D.}\ \bibnamefont
  {Sorkin}},\ }\href@noop {} {\bibfield  {journal} {\bibinfo  {journal}
  {arXiv:gr-qc/9302018}\ } (\bibinfo {year} {1993})}\BibitemShut {NoStop}%
\bibitem [{\citenamefont {Dowker}(2011)}]{Dowker}%
  \BibitemOpen
  \bibfield  {author} {\bibinfo {author} {\bibfnamefont {F.}~\bibnamefont
  {Dowker}},\ }\href@noop {} {\bibfield  {journal} {\bibinfo  {journal}
  {arXiv:1111.2308}\ } (\bibinfo {year} {2011})}\BibitemShut {NoStop}%
\bibitem [{\citenamefont {Benincasa}\ \emph {et~al.}(2014)\citenamefont
  {Benincasa}, \citenamefont {Borsten}, \citenamefont {Buck},\ and\
  \citenamefont {Dowker}}]{Dowker2}%
  \BibitemOpen
  \bibfield  {author} {\bibinfo {author} {\bibfnamefont {D.~M.~T.}\
  \bibnamefont {Benincasa}}, \bibinfo {author} {\bibfnamefont {L.}~\bibnamefont
  {Borsten}}, \bibinfo {author} {\bibfnamefont {M.}~\bibnamefont {Buck}}, \
  and\ \bibinfo {author} {\bibfnamefont {F.}~\bibnamefont {Dowker}},\ }\href
  {\doibase 10.1088/0264-9381/31/7/075007} {\bibfield  {journal} {\bibinfo
  {journal} {Class. Quantum Grav.}\ }\textbf {\bibinfo {volume} {31}},\
  \bibinfo {pages} {075007} (\bibinfo {year} {2014})}\BibitemShut {NoStop}%
\bibitem [{\citenamefont {Christopher J.~Fewster}(2018)}]{Fewster}%
  \BibitemOpen
  \bibfield  {author} {\bibinfo {author} {\bibfnamefont {R.~V.}\ \bibnamefont
  {Christopher J.~Fewster}},\ }\href@noop {} {\bibfield  {journal} {\bibinfo
  {journal} {arXiv:1810.06512}\ } (\bibinfo {year} {2018})}\BibitemShut
  {NoStop}%
\bibitem [{\citenamefont {Mart\'{\i}n-Mart\'{\i}nez}\ and\ \citenamefont
  {Rodriguez-Lopez}(2018)}]{Martin-Martinez2018}%
  \BibitemOpen
  \bibfield  {author} {\bibinfo {author} {\bibfnamefont {E.}~\bibnamefont
  {Mart\'{\i}n-Mart\'{\i}nez}}\ and\ \bibinfo {author} {\bibfnamefont
  {P.}~\bibnamefont {Rodriguez-Lopez}},\ }\href {\doibase
  10.1103/PhysRevD.97.105026} {\bibfield  {journal} {\bibinfo  {journal} {Phys.
  Rev. D}\ }\textbf {\bibinfo {volume} {97}},\ \bibinfo {pages} {105026}
  (\bibinfo {year} {2018})}\BibitemShut {NoStop}%
\bibitem [{\citenamefont {Bartolotta}\ and\ \citenamefont
  {Deffner}(2018)}]{bartolotta2018jarzynski}%
  \BibitemOpen
  \bibfield  {author} {\bibinfo {author} {\bibfnamefont {A.}~\bibnamefont
  {Bartolotta}}\ and\ \bibinfo {author} {\bibfnamefont {S.}~\bibnamefont
  {Deffner}},\ }\href@noop {} {\bibfield  {journal} {\bibinfo  {journal} {Phys.
  Rev. X}\ }\textbf {\bibinfo {volume} {8}},\ \bibinfo {pages} {011033}
  (\bibinfo {year} {2018})}\BibitemShut {NoStop}%
\bibitem [{\citenamefont {Dorner}\ \emph {et~al.}(2013)\citenamefont {Dorner},
  \citenamefont {Clark}, \citenamefont {Heaney}, \citenamefont {Fazio},
  \citenamefont {Goold},\ and\ \citenamefont {Vedral}}]{Dorner2013}%
  \BibitemOpen
  \bibfield  {author} {\bibinfo {author} {\bibfnamefont {R.}~\bibnamefont
  {Dorner}}, \bibinfo {author} {\bibfnamefont {S.~R.}\ \bibnamefont {Clark}},
  \bibinfo {author} {\bibfnamefont {L.}~\bibnamefont {Heaney}}, \bibinfo
  {author} {\bibfnamefont {R.}~\bibnamefont {Fazio}}, \bibinfo {author}
  {\bibfnamefont {J.}~\bibnamefont {Goold}}, \ and\ \bibinfo {author}
  {\bibfnamefont {V.}~\bibnamefont {Vedral}},\ }\href {\doibase
  10.1103/PhysRevLett.110.230601} {\bibfield  {journal} {\bibinfo  {journal}
  {Phys. Rev. Lett.}\ }\textbf {\bibinfo {volume} {110}},\ \bibinfo {pages}
  {230601} (\bibinfo {year} {2013})}\BibitemShut {NoStop}%
\bibitem [{\citenamefont {Mazzola}\ \emph {et~al.}(2014)\citenamefont
  {Mazzola}, \citenamefont {Chiara},\ and\ \citenamefont
  {Paternostro}}]{Mazzola2014}%
  \BibitemOpen
  \bibfield  {author} {\bibinfo {author} {\bibfnamefont {L.}~\bibnamefont
  {Mazzola}}, \bibinfo {author} {\bibfnamefont {G.~D.}\ \bibnamefont {Chiara}},
  \ and\ \bibinfo {author} {\bibfnamefont {M.}~\bibnamefont {Paternostro}},\
  }\href {\doibase 10.1142/S0219749914610073} {\bibfield  {journal} {\bibinfo
  {journal} {Int. J. Quantum Inf.}\ }\textbf {\bibinfo {volume} {12}},\
  \bibinfo {pages} {1461007} (\bibinfo {year} {2014})}\BibitemShut {NoStop}%
\bibitem [{\citenamefont {Batalh\~ao}\ \emph {et~al.}(2014)\citenamefont
  {Batalh\~ao}, \citenamefont {Souza}, \citenamefont {Mazzola}, \citenamefont
  {Auccaise}, \citenamefont {Sarthour}, \citenamefont {Oliveira}, \citenamefont
  {Goold}, \citenamefont {De~Chiara}, \citenamefont {Paternostro},\ and\
  \citenamefont {Serra}}]{batalhaoexp}%
  \BibitemOpen
  \bibfield  {author} {\bibinfo {author} {\bibfnamefont {T.~B.}\ \bibnamefont
  {Batalh\~ao}}, \bibinfo {author} {\bibfnamefont {A.~M.}\ \bibnamefont
  {Souza}}, \bibinfo {author} {\bibfnamefont {L.}~\bibnamefont {Mazzola}},
  \bibinfo {author} {\bibfnamefont {R.}~\bibnamefont {Auccaise}}, \bibinfo
  {author} {\bibfnamefont {R.~S.}\ \bibnamefont {Sarthour}}, \bibinfo {author}
  {\bibfnamefont {I.~S.}\ \bibnamefont {Oliveira}}, \bibinfo {author}
  {\bibfnamefont {J.}~\bibnamefont {Goold}}, \bibinfo {author} {\bibfnamefont
  {G.}~\bibnamefont {De~Chiara}}, \bibinfo {author} {\bibfnamefont
  {M.}~\bibnamefont {Paternostro}}, \ and\ \bibinfo {author} {\bibfnamefont
  {R.~M.}\ \bibnamefont {Serra}},\ }\href {\doibase
  10.1103/PhysRevLett.113.140601} {\bibfield  {journal} {\bibinfo  {journal}
  {Phys. Rev. Lett.}\ }\textbf {\bibinfo {volume} {113}},\ \bibinfo {pages}
  {140601} (\bibinfo {year} {2014})}\BibitemShut {NoStop}%
\bibitem [{\citenamefont {Kubo}(1957)}]{Kubo}%
  \BibitemOpen
  \bibfield  {author} {\bibinfo {author} {\bibfnamefont {R.}~\bibnamefont
  {Kubo}},\ }\href {\doibase 10.1143/JPSJ.12.570} {\bibfield  {journal}
  {\bibinfo  {journal} {J. Phys. Soc. Jpn}\ }\textbf {\bibinfo {volume} {12}},\
  \bibinfo {pages} {570} (\bibinfo {year} {1957})}\BibitemShut {NoStop}%
\bibitem [{\citenamefont {Martin}\ and\ \citenamefont
  {Schwinger}(1959)}]{schwinger}%
  \BibitemOpen
  \bibfield  {author} {\bibinfo {author} {\bibfnamefont {P.~C.}\ \bibnamefont
  {Martin}}\ and\ \bibinfo {author} {\bibfnamefont {J.}~\bibnamefont
  {Schwinger}},\ }\href {\doibase 10.1103/PhysRev.115.1342} {\bibfield
  {journal} {\bibinfo  {journal} {Phys. Rev.}\ }\textbf {\bibinfo {volume}
  {115}},\ \bibinfo {pages} {1342} (\bibinfo {year} {1959})}\BibitemShut
  {NoStop}%
\bibitem [{\citenamefont {Luigi~Accardi}\ and\ \citenamefont
  {Volovich}(2002)}]{kms}%
  \BibitemOpen
  \bibfield  {author} {\bibinfo {author} {\bibfnamefont {Y.~G.~L.}\
  \bibnamefont {Luigi~Accardi}}\ and\ \bibinfo {author} {\bibfnamefont
  {I.}~\bibnamefont {Volovich}},\ }\href@noop {} {\emph {\bibinfo {title}
  {Quantum Theory an its Stochastic Limit}}}\ (\bibinfo  {publisher}
  {Springer},\ \bibinfo {year} {2002})\BibitemShut {NoStop}%
\bibitem [{\citenamefont {Garay}\ \emph {et~al.}(2016)\citenamefont {Garay},
  \citenamefont {Mart\'{\i}n-Mart\'{\i}nez},\ and\ \citenamefont
  {de~Ram\'on}}]{Pipo}%
  \BibitemOpen
  \bibfield  {author} {\bibinfo {author} {\bibfnamefont {L.~J.}\ \bibnamefont
  {Garay}}, \bibinfo {author} {\bibfnamefont {E.}~\bibnamefont
  {Mart\'{\i}n-Mart\'{\i}nez}}, \ and\ \bibinfo {author} {\bibfnamefont
  {J.}~\bibnamefont {de~Ram\'on}},\ }\href {\doibase
  10.1103/PhysRevD.94.104048} {\bibfield  {journal} {\bibinfo  {journal} {Phys.
  Rev. D}\ }\textbf {\bibinfo {volume} {94}},\ \bibinfo {pages} {104048}
  (\bibinfo {year} {2016})}\BibitemShut {NoStop}%
\bibitem [{\citenamefont {Scully}\ and\ \citenamefont
  {Zubairy}(1997)}]{ScullyBook}%
  \BibitemOpen
  \bibfield  {author} {\bibinfo {author} {\bibfnamefont {M.~O.}\ \bibnamefont
  {Scully}}\ and\ \bibinfo {author} {\bibfnamefont {M.~S.}\ \bibnamefont
  {Zubairy}},\ }\href@noop {} {\emph {\bibinfo {title} {Quantum Optics}}}\
  (\bibinfo  {publisher} {Cambridge University Press},\ \bibinfo {year}
  {1997})\BibitemShut {NoStop}%
\bibitem [{\citenamefont {Simidzija}\ and\ \citenamefont
  {Mart\'{\i}n-Mart\'{\i}nez}(2018)}]{petar}%
  \BibitemOpen
  \bibfield  {author} {\bibinfo {author} {\bibfnamefont {P.}~\bibnamefont
  {Simidzija}}\ and\ \bibinfo {author} {\bibfnamefont {E.}~\bibnamefont
  {Mart\'{\i}n-Mart\'{\i}nez}},\ }\href {\doibase 10.1103/PhysRevD.98.085007}
  {\bibfield  {journal} {\bibinfo  {journal} {Phys. Rev. D}\ }\textbf {\bibinfo
  {volume} {98}},\ \bibinfo {pages} {085007} (\bibinfo {year}
  {2018})}\BibitemShut {NoStop}%
\bibitem [{\citenamefont {Crooks}(1999)}]{crooks1999entropy}%
  \BibitemOpen
  \bibfield  {author} {\bibinfo {author} {\bibfnamefont {G.~E.}\ \bibnamefont
  {Crooks}},\ }\href@noop {} {\bibfield  {journal} {\bibinfo  {journal} {Phys.
  Rev. E}\ }\textbf {\bibinfo {volume} {60}},\ \bibinfo {pages} {2721}
  (\bibinfo {year} {1999})}\BibitemShut {NoStop}%
\bibitem [{\citenamefont {Simidzija}\ and\ \citenamefont
  {Mart\'{\i}n-Mart\'{\i}nez}(2017)}]{nonpert}%
  \BibitemOpen
  \bibfield  {author} {\bibinfo {author} {\bibfnamefont {P.}~\bibnamefont
  {Simidzija}}\ and\ \bibinfo {author} {\bibfnamefont {E.}~\bibnamefont
  {Mart\'{\i}n-Mart\'{\i}nez}},\ }\href {\doibase 10.1103/PhysRevD.96.065008}
  {\bibfield  {journal} {\bibinfo  {journal} {Phys. Rev. D}\ }\textbf {\bibinfo
  {volume} {96}},\ \bibinfo {pages} {065008} (\bibinfo {year}
  {2017})}\BibitemShut {NoStop}%
\bibitem [{\citenamefont {Abramowitz}\ and\ \citenamefont
  {Stegun}(1972)}]{Abramowitz}%
  \BibitemOpen
  \bibfield  {author} {\bibinfo {author} {\bibfnamefont {M.}~\bibnamefont
  {Abramowitz}}\ and\ \bibinfo {author} {\bibfnamefont {I.~A.}\ \bibnamefont
  {Stegun}},\ }\href@noop {} {\emph {\bibinfo {title} {Handbook of Mathematical
  Functions with Formulas, Graphs, and Mathematical Tables, 9th printing.}}}\
  (\bibinfo  {publisher} {New York: Dover, pp. 295 and 319},\ \bibinfo {year}
  {1972})\BibitemShut {NoStop}%
\bibitem [{\citenamefont {Albash}\ \emph {et~al.}(2013)\citenamefont {Albash},
  \citenamefont {Lidar}, \citenamefont {Marvian},\ and\ \citenamefont
  {Zanardi}}]{Albash13}%
  \BibitemOpen
  \bibfield  {author} {\bibinfo {author} {\bibfnamefont {T.}~\bibnamefont
  {Albash}}, \bibinfo {author} {\bibfnamefont {D.~A.}\ \bibnamefont {Lidar}},
  \bibinfo {author} {\bibfnamefont {M.}~\bibnamefont {Marvian}}, \ and\
  \bibinfo {author} {\bibfnamefont {P.}~\bibnamefont {Zanardi}},\ }\href
  {\doibase 10.1103/PhysRevE.88.032146} {\bibfield  {journal} {\bibinfo
  {journal} {Phys. Rev. E}\ }\textbf {\bibinfo {volume} {88}},\ \bibinfo
  {pages} {032146} (\bibinfo {year} {2013})}\BibitemShut {NoStop}%
\bibitem [{\citenamefont {Manuceau}\ and\ \citenamefont
  {Verbeure}(1968)}]{algebra}%
  \BibitemOpen
  \bibfield  {author} {\bibinfo {author} {\bibfnamefont {J.}~\bibnamefont
  {Manuceau}}\ and\ \bibinfo {author} {\bibfnamefont {A.}~\bibnamefont
  {Verbeure}},\ }\href {https://projecteuclid.org:443/euclid.cmp/1103840803}
  {\bibfield  {journal} {\bibinfo  {journal} {Comm. Math. Phys.}\ }\textbf
  {\bibinfo {volume} {9}},\ \bibinfo {pages} {293} (\bibinfo {year}
  {1968})}\BibitemShut {NoStop}%
\bibitem [{\citenamefont {Forn-Díaz}\ \emph {et~al.}(2017)\citenamefont
  {Forn-Díaz}, \citenamefont {García-Ripoll}, \citenamefont {Peropadre},
  \citenamefont {Orgiazzi}, \citenamefont {Yurtalan}, \citenamefont
  {Belyansky}, \citenamefont {Wilson},\ and\ \citenamefont
  {Lupascu}}]{experiment}%
  \BibitemOpen
  \bibfield  {author} {\bibinfo {author} {\bibfnamefont {P.}~\bibnamefont
  {Forn-Díaz}}, \bibinfo {author} {\bibfnamefont {J.~J.}\ \bibnamefont
  {García-Ripoll}}, \bibinfo {author} {\bibfnamefont {B.}~\bibnamefont
  {Peropadre}}, \bibinfo {author} {\bibfnamefont {J.-L.}\ \bibnamefont
  {Orgiazzi}}, \bibinfo {author} {\bibfnamefont {M.~A.}\ \bibnamefont
  {Yurtalan}}, \bibinfo {author} {\bibfnamefont {R.}~\bibnamefont {Belyansky}},
  \bibinfo {author} {\bibfnamefont {C.~M.}\ \bibnamefont {Wilson}}, \ and\
  \bibinfo {author} {\bibfnamefont {A.}~\bibnamefont {Lupascu}},\ }\href@noop
  {} {\bibfield  {journal} {\bibinfo  {journal} {Nature}\ }\textbf {\bibinfo
  {volume} {13}},\ \bibinfo {pages} {39} (\bibinfo {year} {2017})}\BibitemShut
  {NoStop}%
\bibitem [{\citenamefont {Campisi}\ \emph {et~al.}(2013)\citenamefont
  {Campisi}, \citenamefont {Blattmann}, \citenamefont {Kohler}, \citenamefont
  {Zueco},\ and\ \citenamefont {Hänggi}}]{circuit}%
  \BibitemOpen
  \bibfield  {author} {\bibinfo {author} {\bibfnamefont {M.}~\bibnamefont
  {Campisi}}, \bibinfo {author} {\bibfnamefont {R.}~\bibnamefont {Blattmann}},
  \bibinfo {author} {\bibfnamefont {S.}~\bibnamefont {Kohler}}, \bibinfo
  {author} {\bibfnamefont {D.}~\bibnamefont {Zueco}}, \ and\ \bibinfo {author}
  {\bibfnamefont {P.}~\bibnamefont {Hänggi}},\ }\href {\doibase
  10.1088/1367-2630/15/10/105028} {\bibfield  {journal} {\bibinfo  {journal}
  {New J. Phys.}\ }\textbf {\bibinfo {volume} {15}},\ \bibinfo {pages} {105028}
  (\bibinfo {year} {2013})}\BibitemShut {NoStop}%
\bibitem [{\citenamefont {Sab\'{\i}n}\ \emph {et~al.}(2012)\citenamefont
  {Sab\'{\i}n}, \citenamefont {Peropadre}, \citenamefont {del Rey},\ and\
  \citenamefont {Mart\'{\i}n-Mart\'{\i}nez}}]{pastfuture}%
  \BibitemOpen
  \bibfield  {author} {\bibinfo {author} {\bibfnamefont {C.}~\bibnamefont
  {Sab\'{\i}n}}, \bibinfo {author} {\bibfnamefont {B.}~\bibnamefont
  {Peropadre}}, \bibinfo {author} {\bibfnamefont {M.}~\bibnamefont {del Rey}},
  \ and\ \bibinfo {author} {\bibfnamefont {E.}~\bibnamefont
  {Mart\'{\i}n-Mart\'{\i}nez}},\ }\href {\doibase
  10.1103/PhysRevLett.109.033602} {\bibfield  {journal} {\bibinfo  {journal}
  {Phys. Rev. Lett.}\ }\textbf {\bibinfo {volume} {109}},\ \bibinfo {pages}
  {033602} (\bibinfo {year} {2012})}\BibitemShut {NoStop}%
\bibitem [{\citenamefont {Garc\'{i}a-\'{A}lvarez}\ \emph
  {et~al.}(2017)\citenamefont {Garc\'{i}a-\'{A}lvarez}, \citenamefont
  {Felicetti}, \citenamefont {Rico}, \citenamefont {Solano},\ and\
  \citenamefont {Sab\'{i}n}}]{entanglementacceleration}%
  \BibitemOpen
  \bibfield  {author} {\bibinfo {author} {\bibfnamefont {L.}~\bibnamefont
  {Garc\'{i}a-\'{A}lvarez}}, \bibinfo {author} {\bibfnamefont {S.}~\bibnamefont
  {Felicetti}}, \bibinfo {author} {\bibfnamefont {E.}~\bibnamefont {Rico}},
  \bibinfo {author} {\bibfnamefont {E.}~\bibnamefont {Solano}}, \ and\ \bibinfo
  {author} {\bibfnamefont {C.}~\bibnamefont {Sab\'{i}n}},\ }\href@noop {}
  {\bibfield  {journal} {\bibinfo  {journal} {Sci. Rep.}\ }\textbf {\bibinfo
  {volume} {7}} (\bibinfo {year} {2017})}\BibitemShut {NoStop}%
\bibitem [{\citenamefont {Garc\'{\i}a-\'Alvarez}\ \emph
  {et~al.}(2015)\citenamefont {Garc\'{\i}a-\'Alvarez}, \citenamefont
  {Casanova}, \citenamefont {Mezzacapo}, \citenamefont {Egusquiza},
  \citenamefont {Lamata}, \citenamefont {Romero},\ and\ \citenamefont
  {Solano}}]{fermion}%
  \BibitemOpen
  \bibfield  {author} {\bibinfo {author} {\bibfnamefont {L.}~\bibnamefont
  {Garc\'{\i}a-\'Alvarez}}, \bibinfo {author} {\bibfnamefont {J.}~\bibnamefont
  {Casanova}}, \bibinfo {author} {\bibfnamefont {A.}~\bibnamefont {Mezzacapo}},
  \bibinfo {author} {\bibfnamefont {I.~L.}\ \bibnamefont {Egusquiza}}, \bibinfo
  {author} {\bibfnamefont {L.}~\bibnamefont {Lamata}}, \bibinfo {author}
  {\bibfnamefont {G.}~\bibnamefont {Romero}}, \ and\ \bibinfo {author}
  {\bibfnamefont {E.}~\bibnamefont {Solano}},\ }\href {\doibase
  10.1103/PhysRevLett.114.070502} {\bibfield  {journal} {\bibinfo  {journal}
  {Phys. Rev. Lett.}\ }\textbf {\bibinfo {volume} {114}},\ \bibinfo {pages}
  {070502} (\bibinfo {year} {2015})}\BibitemShut {NoStop}%
\bibitem [{\citenamefont {Liphardt}\ \emph {et~al.}(2002)\citenamefont
  {Liphardt}, \citenamefont {Dumont}, \citenamefont {Smith}, \citenamefont
  {Tinoco},\ and\ \citenamefont {Bustamante}}]{liphardt2002equilibrium}%
  \BibitemOpen
  \bibfield  {author} {\bibinfo {author} {\bibfnamefont {J.}~\bibnamefont
  {Liphardt}}, \bibinfo {author} {\bibfnamefont {S.}~\bibnamefont {Dumont}},
  \bibinfo {author} {\bibfnamefont {S.~B.}\ \bibnamefont {Smith}}, \bibinfo
  {author} {\bibfnamefont {I.}~\bibnamefont {Tinoco}}, \ and\ \bibinfo {author}
  {\bibfnamefont {C.}~\bibnamefont {Bustamante}},\ }\href@noop {} {\bibfield
  {journal} {\bibinfo  {journal} {Science}\ }\textbf {\bibinfo {volume}
  {296}},\ \bibinfo {pages} {1832} (\bibinfo {year} {2002})}\BibitemShut
  {NoStop}%
\bibitem [{\citenamefont {Park}\ \emph {et~al.}(2003)\citenamefont {Park},
  \citenamefont {Khalili-Araghi}, \citenamefont {Tajkhorshid},\ and\
  \citenamefont {Schulten}}]{park2003free}%
  \BibitemOpen
  \bibfield  {author} {\bibinfo {author} {\bibfnamefont {S.}~\bibnamefont
  {Park}}, \bibinfo {author} {\bibfnamefont {F.}~\bibnamefont
  {Khalili-Araghi}}, \bibinfo {author} {\bibfnamefont {E.}~\bibnamefont
  {Tajkhorshid}}, \ and\ \bibinfo {author} {\bibfnamefont {K.}~\bibnamefont
  {Schulten}},\ }\href@noop {} {\bibfield  {journal} {\bibinfo  {journal} {J.
  Chem. Phys.}\ }\textbf {\bibinfo {volume} {119}},\ \bibinfo {pages} {3559}
  (\bibinfo {year} {2003})}\BibitemShut {NoStop}%
\bibitem [{\citenamefont {Collin}\ \emph {et~al.}(2005)\citenamefont {Collin},
  \citenamefont {Ritort}, \citenamefont {Jarzynski}, \citenamefont {Smith},
  \citenamefont {Tinoco~Jr},\ and\ \citenamefont
  {Bustamante}}]{collin2005verification}%
  \BibitemOpen
  \bibfield  {author} {\bibinfo {author} {\bibfnamefont {D.}~\bibnamefont
  {Collin}}, \bibinfo {author} {\bibfnamefont {F.}~\bibnamefont {Ritort}},
  \bibinfo {author} {\bibfnamefont {C.}~\bibnamefont {Jarzynski}}, \bibinfo
  {author} {\bibfnamefont {S.~B.}\ \bibnamefont {Smith}}, \bibinfo {author}
  {\bibfnamefont {I.}~\bibnamefont {Tinoco~Jr}}, \ and\ \bibinfo {author}
  {\bibfnamefont {C.}~\bibnamefont {Bustamante}},\ }\href@noop {} {\bibfield
  {journal} {\bibinfo  {journal} {Nature}\ }\textbf {\bibinfo {volume} {437}},\
  \bibinfo {pages} {231} (\bibinfo {year} {2005})}\BibitemShut {NoStop}%
\bibitem [{\citenamefont {Pozas-Kerstjens}\ and\ \citenamefont
  {Mart\'{\i}n-Mart\'{\i}nez}(2015)}]{entanglealex}%
  \BibitemOpen
  \bibfield  {author} {\bibinfo {author} {\bibfnamefont {A.}~\bibnamefont
  {Pozas-Kerstjens}}\ and\ \bibinfo {author} {\bibfnamefont {E.}~\bibnamefont
  {Mart\'{\i}n-Mart\'{\i}nez}},\ }\href {\doibase 10.1103/PhysRevD.92.064042}
  {\bibfield  {journal} {\bibinfo  {journal} {Phys. Rev. D}\ }\textbf {\bibinfo
  {volume} {92}},\ \bibinfo {pages} {064042} (\bibinfo {year}
  {2015})}\BibitemShut {NoStop}%
\bibitem [{\citenamefont {Summers}\ and\ \citenamefont
  {Werner}(1985)}]{entangbell}%
  \BibitemOpen
  \bibfield  {author} {\bibinfo {author} {\bibfnamefont {S.~J.}\ \bibnamefont
  {Summers}}\ and\ \bibinfo {author} {\bibfnamefont {R.}~\bibnamefont
  {Werner}},\ }\href {\doibase https://doi.org/10.1016/0375-9601(85)90093-3}
  {\bibfield  {journal} {\bibinfo  {journal} {Phys. Lett.}\ }\textbf {\bibinfo
  {volume} {110}},\ \bibinfo {pages} {257} (\bibinfo {year}
  {1985})}\BibitemShut {NoStop}%
\bibitem [{\citenamefont {Reznik}(2003)}]{entvac}%
  \BibitemOpen
  \bibfield  {author} {\bibinfo {author} {\bibfnamefont {B.}~\bibnamefont
  {Reznik}},\ }\href {\doibase 10.1023/A:1022875910744} {\bibfield  {journal}
  {\bibinfo  {journal} {Found. Phys.}\ }\textbf {\bibinfo {volume} {33}},\
  \bibinfo {pages} {167} (\bibinfo {year} {2003})}\BibitemShut {NoStop}%
\bibitem [{\citenamefont {Valentini}(1991)}]{nonlocal}%
  \BibitemOpen
  \bibfield  {author} {\bibinfo {author} {\bibfnamefont {A.}~\bibnamefont
  {Valentini}},\ }\href {\doibase https://doi.org/10.1016/0375-9601(91)90952-5}
  {\bibfield  {journal} {\bibinfo  {journal} {Phys. Lett.}\ }\textbf {\bibinfo
  {volume} {153}},\ \bibinfo {pages} {321} (\bibinfo {year}
  {1991})}\BibitemShut {NoStop}%
\bibitem [{\citenamefont {Hotta}(2008)}]{qet}%
  \BibitemOpen
  \bibfield  {author} {\bibinfo {author} {\bibfnamefont {M.}~\bibnamefont
  {Hotta}},\ }\href {\doibase 10.1103/PhysRevD.78.045006} {\bibfield  {journal}
  {\bibinfo  {journal} {Phys. Rev. D}\ }\textbf {\bibinfo {volume} {78}},\
  \bibinfo {pages} {045006} (\bibinfo {year} {2008})}\BibitemShut {NoStop}%
\bibitem [{\citenamefont {Fewster}\ \emph {et~al.}(2016)\citenamefont
  {Fewster}, \citenamefont {Ju{\'{a}}rez-Aubry},\ and\ \citenamefont
  {Louko}}]{waiting}%
  \BibitemOpen
  \bibfield  {author} {\bibinfo {author} {\bibfnamefont {C.~J.}\ \bibnamefont
  {Fewster}}, \bibinfo {author} {\bibfnamefont {B.~A.}\ \bibnamefont
  {Ju{\'{a}}rez-Aubry}}, \ and\ \bibinfo {author} {\bibfnamefont
  {J.}~\bibnamefont {Louko}},\ }\href {\doibase 10.1088/0264-9381/33/16/165003}
  {\bibfield  {journal} {\bibinfo  {journal} {Class. Quantum Gravity}\ }\textbf
  {\bibinfo {volume} {33}},\ \bibinfo {pages} {165003} (\bibinfo {year}
  {2016})}\BibitemShut {NoStop}%
\end{thebibliography}%

\end{document}